\documentclass[letterpaper, 10 pt, onecolumn]{ieeeconf} 

\IEEEoverridecommandlockouts                              % This command is only needed if 
\usepackage{graphicx} 
\usepackage{amsmath} % assumes amsmath package installed
\usepackage{amssymb}  % assumes amsmath package installed
\usepackage{mathrsfs}% for cursive letters in math formulas
\usepackage{ulem}
\usepackage[usenames]{color}
\newcounter{theorem}
\newcommand{\theorem}{\refstepcounter{theorem}{\bf Theorem
\arabic{theorem}. }}
\newcounter{lemma}
\newcommand{\lemma}{\refstepcounter{lemma}{\bf Lemma
\arabic{lemma}. }}
\newcounter{corollary}
\newcommand{\corollary}{\refstepcounter{corollary}{\bf Corollary
\arabic{corollary}. }}
\newcounter{remark}
\newcommand{\remark}{\refstepcounter{remark}{\bf Remark
\arabic{remark}. }}
\newcounter{definition}
\newcommand{\definition}{\refstepcounter{definition}{\bf Definition
\arabic{definition}. }}
\newcounter{assumption}
\newcommand{\assumption}{\refstepcounter{assumption}{\bf Assumption
\arabic{assumption}. }}
\newcounter{property}

\newcounter{condition}

\newcounter{proc}
\newcommand{\procedure}{\refstepcounter{proc}{\bf Procedure
\arabic{proc}. }}
\title{\LARGE \bf
Interconnections of Dissipative Networks Through a Dynamic Scattering Controller
}

\author{Ilia G. Polushin$^{1}$% <-this % stops a space
\thanks{*The research was supported by the Discovery Grants Program of the Natural Sciences and Engineering Research Council (NSERC) of Canada through grants RGPIN-2020-06441.}% <-this % stops a space
\thanks{$^{1}$The author is with the Department of Electrical and Computer Engineering, Western University, London, ON, N6A 5B9, Canada. Email: {\tt\small ipolushi@uwo.ca}}%
}

\begin{document}

\maketitle
\thispagestyle{empty}
\pagestyle{empty}

%%%%%%%%%%%%%%%%%%%%%%%%%%%%%%%%%%%%%%%%%%%%%%%%%%%%%%%%%%%%%%%%%%%%%%%%%%%%%%%%
\begin{abstract} The problem of modular scattering-based design of dissipative networks is addressed. Under basic assumptions imposed on subsystems, design of a dynamic scattering-based interconnection controller is proposed which guarantees that the resulting interconnection is internally stable and possesses the same set of basic properties as the individual subsystems. This enables iterative application of the developed technique, which allows for building of large-scale dissipative networks in a modular fashion. The proposed method provides substantial flexibility in the controller design as well as in the choice of input and output signals used for interconnections. 
\end{abstract}

\section{Introduction} The theory of dissipative systems is a well-established research area with a number of important applications to both theoretical and practical problems related to systems and control~\cite{brogliato:lozano:maschke:egeland:20}. In particular, it was suggested as early as the 1970s that dissipative systems theory provides a natural framework for stability analysis of large scale interconnections~\cite{moylan:hill:1978:largescale}. One of the popular early results related to interconnections of dissipative systems is the so-called passivity theorem~\cite{desoer:vidyasagar:75}, which states that the feedback interconnection of two passive systems is itself a passive system. This result stimulated interest in the passivation (or passification) problem, namely, the problem of achieving passivity by combination of state/output feedback and output assignment~\cite{byrnes:isidori:willems:91,sepulchre:jankovic:kokotovic:97,fradkov:hill:1998:automatica,Polushin:Fradkov:Hill:AiT:2000}. From dissipativity point of view, passive systems are dissipative with a special type of quadratic supply rate. Stability results for interconnections of dissipative systems with more general supply rates are currently abundant~\cite{moylan:hill:1978:largescale,Willems:Takaba:2007}, \cite[Chapter 2]{Arcak:Meissen:Packard:2016:book}, \cite{Polushin:tac:2014,usova:polushin:patel:18:automatica,hill:liu:2022:csm,scherer:ieeecsm:2022}. Generally speaking, an interconnection of dissipative systems is stable if the corresponding supply rates satisfy some form of graph separation condition. Scattering transformations~\cite{anderson:spong:89,Hirche:etal:2009:aut,Polushin:tac:2014,usova:polushin:patel:18:automatica,usova:polushin:patel:19:cssletters} provide a tool for assigning supply rate functions in dissipative systems. A systematic approach to the scattering-based design of large scale dissipative networks was recently developed in~\cite{polushin:26:automatica}, see also~\cite{polushin:25:cdc} for the case of communication delays. In particular, in~\cite{polushin:26:automatica}, scattering transformations are used for interconnecting the subsystems so that the resulting interconnection is itself a dissipative network with a well defined supply rate, and therefore can be further interconnected with similar networks using the same method. Systematic application of this procedure leads to modular design of complex dissipative networks.

In this paper, the problem of modular design of large-scale dissipative networks through appropriately designed dynamic scattering-based interconnection controllers is addressed. 
The results presented in this paper improve upon the results in~\cite{polushin:26:automatica} in several fundamental aspects. One major advantage of the method presented in this paper over that of~\cite{polushin:26:automatica} is that it guarantees that the supply rate of the interconnection satisfies exactly the same conditions as those imposed on the subsystems, which in turn guarantees the possibility of building arbitrarily large networks out of such dissipative components. In contrast, the method presented in~\cite{polushin:26:automatica} does not generally allow for control over the supply rate of interconnection and, moreover, does not guarantee that the supply rate of the interconnection satisfies the set of conditions that allow for its use as a building block in a larger interconnection.   In other words, conditions imposed on subsystems that allow for their stable interconnection in~\cite{polushin:26:automatica} are not guaranteed to be satisfied by interconnection itself. In fact, the method of calculating the supply rate of an interconnection presented in~\cite{polushin:26:automatica} uses approximations that typically leads to supply rates estimates for which the liveness condition eventually fails to be satisfied. As a result, such an interconnection cannot be further interconnected with other dissipative subsystems, which essentially puts a limit on the scale of networks that can be built using the method of~\cite{polushin:26:automatica}.  Other advantages of the technique developed in this paper is that it allows for interconnection of dissimilar dissipative networks, {\it i.e.}, those with not necessary matching numbers of inputs and outputs, as well as for in some sense complete flexibility in the choice of input-output signals used for interconnection. The controller design process  is also characterized by substantial flexibility which can potentially be used for achieving secondary goals such as performance optimization.

 The paper is organized as follows. Section~\ref{SectionDissipativeSystems01a} presents background information on dissipative systems formulated within the behavioral modeling framework~\cite{willems:polderman:1998:book,willems:2007:ejc}. In Section~\ref{SectionProblemStatementAssumptions01} the problem addressed in this paper is formulated and basic assumptions are introduced. The proposed design of dynamic scattering-based interconnection controller is described in Section~\ref{SectionDesignInterconnectionController01}. Dissipativity properties of the resulting interconnection are studied in detail in Section~\ref{SectionDissipativityPropertiesInterconnection01}. Conclusions are given in Section~\ref{SectionConclusions01}. Appendices~\ref{AppendixProofs01} and~\ref{AppendixRealizability01} contain proofs of certain technical results and discussion related to realizability of scattering transformations, respectively.

%\newpage 

\section{Dissipative systems}\label{SectionDissipativeSystems01a} In this paper, the behavioral approach to mathematical modeling of dissipative systems is used.  Both the notion of dissipative systems and the behavioral approach to systems' modeling are due to Willems, see~\cite{willems:2007:ejc} and references therein. Following behavioral approach, a dynamical system is described by its behavior, which is essentially a collection of trajectories compatible with the dynamics of the system. Let ${\mathbb T}$ be a set of time instants; throughout this paper  ${\mathbb T}={\mathbb R}_+:=[0, +\infty)$. A dynamical system $\Sigma$ interacts with the outside world through a vector of {\it manifest variables} $w\in\mathscr{F}\left({\mathbb R}_+, {\mathbb W}\right)$, where ${\mathbb W}\subset {\mathbb R}^n$ is called {\it signal space}, and $\mathscr{F}\left({\mathbb R}_+, {\mathbb W}\right)$ is the set of continuous locally integrable maps from ${\mathbb R}_+$ to ${\mathbb W}$. The {\it manifest behavior} ${\mathfrak B}(\Sigma)$  is defined as the set of all signals $w\in\mathscr{F}\left({\mathbb R}_+, {\mathbb W}\right)$ compatible with the dynamics of the system $\Sigma$. In the case of input-output systems, which is addressed in this paper, the signal space is partitioned as 
${\mathbb W}:={\mathbb W}_i\times {\mathbb W}_o$, where ${\mathbb W}_{in}\subset {\mathbb R}^m$, ${\mathbb W}_o\subset {\mathbb R}^p$, and the manifest variables are decomposed accordingly as $w:=\left(\eta, y \right)$, $\eta\in\mathscr{F}\left({\mathbb R}_+, {\mathbb W}_{in}\right)$, $y\in\mathscr{F}\left({\mathbb R}_+, {\mathbb W}_o\right)$. Variables $\eta$ are called input variables, and $y$ output variables if the following two properties hold~\cite[Definition 3.3.1]{willems:polderman:1998:book}: i) variables $\eta$ are {\it free}, {\it i.e.}, for any $\eta\in\mathscr{F}\left({\mathbb R}_+, {\mathbb W}_{in}\right)$ there exists $y\in\mathscr{F}\left({\mathbb R}_+, {\mathbb W}_o\right)$ such that $(\eta, y)\in {\mathfrak B}(\Sigma)$, and ii) $\eta$ are maximally free, {\it i.e.}, any subset of manifest variables $w$ which strictly contains $\eta$ is not free. 

The notion of dissipative system within the behavioral framework requires introduction of an additional {\it latent} variable ${\mathcal V} \in\mathscr{F}\left({\mathbb R}_+, {\mathbb R}_+\right)$ which corresponds to a {\it storage function}. The full behavior of an input-output system with manifest variables $w$ and a latent variable ${\mathcal V}$ is formally defined as follow: 
\begin{equation*}{\mathfrak B}_{f}\left(\Sigma\right):=\left\{
\mbox{\parbox{6.5cm}{
$\left(\eta, y, {\mathcal V}\right)\in\mathscr{F}\left({\mathbb R}_+,  {\mathbb W}_{in}\times {\mathbb W}_o\times {\mathbb R}_+\right)$ $\vert$\,
 $\left(\eta, y, {\mathcal V}\right)$ is 
 compatible with dynamics of $\Sigma$}} 
\right\} 
\end{equation*}
The following notion is well known~\cite{willems:2007:ejc}.

\definition An input-output system $\Sigma$ with full behavior ${\mathfrak B}_{f}\left(\Sigma\right)$  is called {\it dissipative} with a {\it supply rate} function $s\colon\, {\mathbb W}_i\times {\mathbb W}_o\to {\mathbb R}$ if for any $\left( \eta, y, {\mathcal V}\right)\in {\mathfrak B}_{f}\left(\Sigma\right)$ the following %dissipation 
inequality 
\vspace{-0.1cm}
\begin{equation}
{\mathcal V}(t_1)-{\mathcal V}(t_0)\le \int_{t_0}^{t_1}s\left(\eta(\tau), y(\tau)  \right) d\tau 
\end{equation}
holds for all $t_0, t_1\in {\mathbb R}_+$, $t_1\ge t_0$. $\square$

Throughout this paper, all supply rate functions are quadratic of the form
\begin{equation}\label{QSRFormula01}
s\left(\eta, y  \right):= 
\begin{bmatrix}
\eta^T & y^T
\end{bmatrix}W
\begin{bmatrix}
\eta\\ y\\
\end{bmatrix},
\end{equation} 
where $W=W^T\in{\mathbb R}^{(m+p)\times(m+p)}$.  

\section{Problem Statement and Assumptions}\label{SectionProblemStatementAssumptions01} 
\begin{figure}[h!]
    \begin{center}
        \includegraphics[scale=0.4]{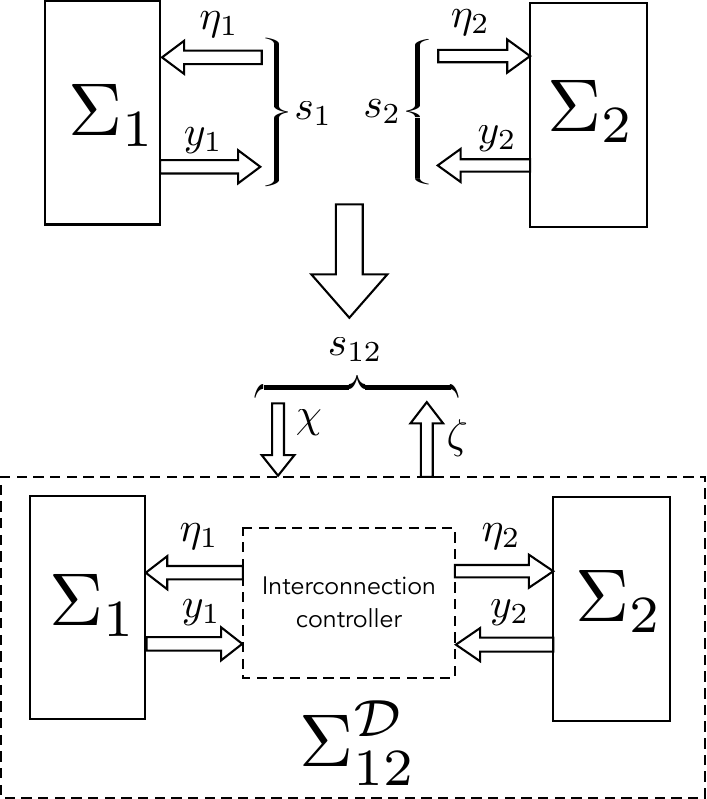}
       \caption{Problem statement}
        \label{Fig:Figure04e}
    \end{center}
\end{figure}

The main problem addressed in this paper is illustrated in Figure~\ref{Fig:Figure04e}. Given two dynamical networks $\Sigma_1$ and $\Sigma_2$ which are dissipative with quadratic supply rates $s_1$ and $s_2$, respectively. %In particular, no constraints are imposed on the numbers of manifest (input and output) variables of each subsystem. 
The matrices of quadratic supply rates $s_1$ and $s_2$ are to satisfy certain technical assumptions. The problem is to design an interconnection controller which  guarantees that the overall interconnection, denoted by $\Sigma_{12}^{\mathcal D}$ in Figure~\ref{Fig:Figure04e},  is internally stable in some appropriate sense and also dissipative with a well-defined quadratic supply rate $s_{12}$, where the latter satisfies the same set of assumptions as those imposed originally on $s_1$ and $s_2$. Thus designed interconnection $\Sigma_{12}^{\mathcal D}$ therefore possesses the same set of essential properties as  each of $\Sigma_1$ and $\Sigma_2$ does and, as a result,  can be further interconnected with other dissipative networks using exactly the same procedure that was used for interconnection of $\Sigma_1$ and $\Sigma_2$. Iterative application of this procedure would therefore allow for step-by-step building of large scale dissipative networks.

For a more rigorous statement of the problem, consider subsystems $\Sigma_1$, $\Sigma_2$ which are described by their full behaviors: 
\begin{equation*}
{\mathfrak B}_{f}\left(\Sigma_i\right)\subset \mathscr{F}\left({\mathbb R}_+,  {\mathbb W}^{\{i\}}_{in}\times {\mathbb W}^{\{i\}}_o\times {\mathbb R}_+\right), \hspace{0.5cm} i=1,2,
\end{equation*}
where the input and the output signal spaces for $i$-th subsystem are denoted by ${\mathbb W}^{\{i\}}_{in}$, ${\mathbb W}^{\{i\}}_{o}$, respectively. For simplicity, throughout this paper we assume ${\mathbb W}^{\{i\}}_{in}= {\mathbb R}^{m_i}$, $ {\mathbb W}^{\{i\}}_o= {\mathbb R}^{p_i}$, where $m_i, p_i\in{\mathbb N}$, $i=1,2$. Thus, subsystem $\Sigma_i$, $i=1,2$, possesses $m_i$ input-type (free) manifest variables, and $p_i$ output-type (dependent) manifest variables; the vectors that comprise these variables are denoted by $\eta_i\in\mathscr{F}\left({\mathbb R}_+, {\mathbb R}^{m_i}\right)$ and $y_i(\cdot)\in\mathscr{F}\left({\mathbb R}_+, {\mathbb R}^{p_i}\right)$, respectively, $i=1,2$. The total number of manifest variables for each $\Sigma_i$ is therefore ${\mathcal N}_i:=m_i+p_i$. 
Note that dimensions of inputs and outputs of $\Sigma_1$ and $\Sigma_2$ are not required to be compatible for direct interconnection which means that, generally speaking, $m_1\ne p_2$ and $m_2\ne p_1$. Subsystems $\Sigma_1$ and $\Sigma_2$ are assumed to satisfy certain basic assumptions, that include dissipativity with a quadratic supply rate, the liveness condition, and realizability condition.  The first assumption can be formulated as follows. 

\assumption\label{AssumptionDissipativity01} Systems $\Sigma_i$ are dissipative with quadratic supply rates of the form
\begin{equation}\label{FormulaSupplyRates01Assumption1}
s_i:= 
\begin{bmatrix}
\eta_i^T & y_i^T
\end{bmatrix}W_i
\begin{bmatrix}
\eta_i\\ y_i
\end{bmatrix}, %\hspace{0.5cm} i=1,2,
\end{equation}
where $W_i=W_i^T\in {\mathbb R}^{{\mathcal N}_i\times {\mathcal N}_i}$, $i=1,2$.  %for $i=1,2$. 
$\bullet$ 

Second assumption is a version of the {\it liveness condition}~\cite{willems:trentelman:part1:tac:2002,usova:polushin:patel:19:cssletters,polushin:26:automatica}, which relates the signature of the quadratic supply rate to the number of free/dependent variables of the system. Below, for a matrix $A=A^T$, $n_-(A)$ denotes the number of negative ($<0$) eigenvalues of $A$. The assumption is formulated as follows. 

\assumption\label{AssumptionLiveness01} The matrices $W_1$, $W_2$ of the supply rates~\eqref{FormulaSupplyRates01Assumption1} satisfy the {\it liveness condition}, {\it i.e.}, $n_{-}\left( W_i\right)=p_i$ for $i=1,2$. $\square$

The final assumption imposed on subsystems $\Sigma_1$, $\Sigma_2$ is that their supply rate matrices satisfy the realizability condition. 
 Specifically, let ${G}^{\{i\}}\in {\mathbb R}^{{\mathcal N}_i\times {\mathcal N}_i}$, $i=1,2$ be orthonormal matrices that satisfy 
\begin{equation*}
W_i{G}^{\{i\}}={G}^{\{i\}}\Lambda_{i}, %\hspace{0.3cm} i=1,2, 
\end{equation*}
where $\Lambda_i$ be diagonal matrices whose main diagonals comprise eigenvalues of $W_i=W_i^T$ written in the descending order. Write ${G}^{\{i\}}$ in the partitioned form, as follows
\begin{equation*}
{G}^{\{i\}}:=\begin{bmatrix}
{G}^{\{i\}}_{11} & {G}^{\{i\}}_{11} \\
{G}^{\{i\}}_{21} & {G}^{\{i\}}_{22} 
\end{bmatrix}, \hspace{0.3cm} i=1,2,
\end{equation*} 
where ${G}^{\{i\}}_{22}\in  {\mathbb R}^{{p}_i\times {p}_i}$. The next and final assumption is formulated as follows. 

\assumption\label{AssumptionRealizability01} ${\mbox{ rank }} {G}^{\{i\}}_{22}=p_i$, $i=1,2$. $\square$

The realizability condition, represented by Assumption~\ref{AssumptionRealizability01}, essentially enables the implementation of the scattering transformation as a feedforward/feedback controller with an output assignment. For a more detailed justification, %of the realizability condition, 
the reader is referred to Appendix~\ref{AppendixRealizability01}. 

\begin{figure}[h!]
    \begin{center}
        \includegraphics[scale=0.4]{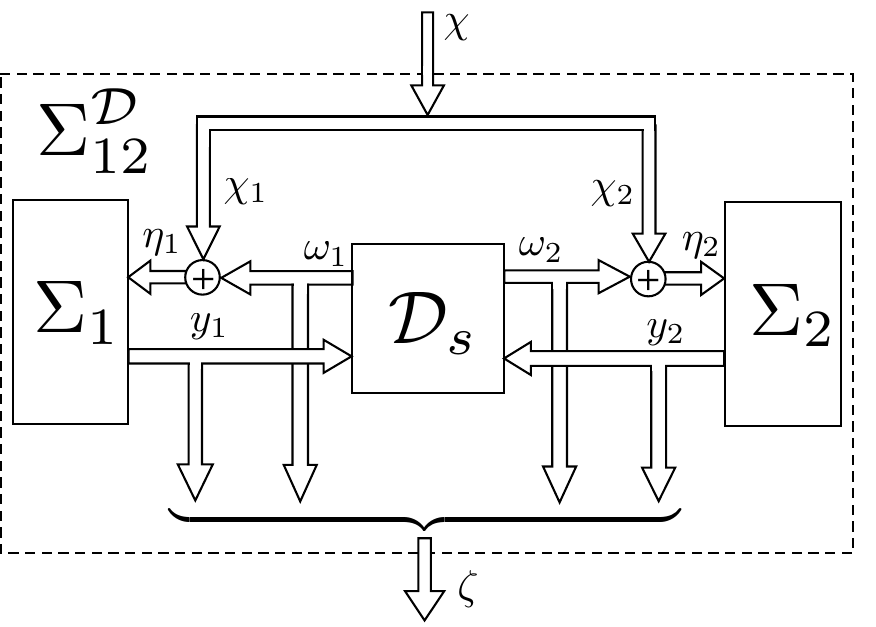}
       \caption{The interconnection structure}
        \label{Fig:Figure04d}
    \end{center}
\end{figure}

Using the above Assumptions, a more detailed formulation of the main problem addressed in this paper can be given as follows. Consider an interconnection structure shown in Figure~\ref{Fig:Figure04d}. 
In this figure, the interconnection $\Sigma_{12}^{\mathcal D}$ comprises subsystems $\Sigma_1$ and $\Sigma_2$ as well as  a scattering-based dynamic controller ${\mathcal D}_s$. The input to $\Sigma_{12}^{\mathcal D}$ is the signal 
\begin{equation}\label{Chi12}
\chi:=\left[\chi^T_1 \, \chi^T_2\right]^T\in\mathscr{F}\left({\mathbb R}_+, {\mathbb R}^{\mathfrak m}\right), 
\end{equation}
where ${\mathfrak m}:=m_1+m_2$. 
The interconnection $\Sigma_{12}^{\mathcal D}$ is formed by the following interconnection constraints 
\begin{equation}\label{InterconnectionConstraints01a}
\eta_i=\chi_i+\omega_i, \hspace{0.5cm}i=1,2
\end{equation}
where $\omega_i$, $i=1,2$, are the outputs of the controller ${\mathcal D}_s$. The output of $\Sigma_{12}^{\mathcal D}$ is the signal 
\begin{equation}\label{Zeta12}
\zeta:=\begin{bmatrix}
\omega_1^T  & y_1^T & \omega_2^T & y_2^T 
\end{bmatrix}^T
\in \mathscr{F}\left({\mathbb R}_+, {\mathbb R}^{{\mathcal N}_1+{\mathcal N}_2}\right),
\end{equation}
which comprises outputs $y_1$, $y_2$ of $\Sigma_1$ and $\Sigma_2$, respectively, and controller outputs $\omega_1$, $\omega_2$. The main problem addressed in this work can be formulated as follows.

{\bf Problem:} Given dynamical networks $\Sigma_1$ and $\Sigma_2$ that satisfy Assumptions~\ref{AssumptionDissipativity01}--\ref{AssumptionRealizability01}, design a (dynamic scattering-based) controller ${\mathcal D}_s$ such that the interconnection $\Sigma_{12}^{\mathcal D}$ with input~\eqref{Chi12} and output~\eqref{Zeta12} has the following properties: 

i) $\Sigma_{12}^{\mathcal D}$ satisfies Assumptions~\ref{AssumptionDissipativity01}--\ref{AssumptionRealizability01}; 

ii) there exists ${\Upsilon}={\Upsilon}^T\in {\mathbb R}^{\mathfrak m}\times {\mathfrak m}$,  ${\Upsilon}\succeq 0$, such that the inequality
\begin{equation}\label{InequalityL2GainOfSigmaD12AA}
\int\limits_{t_0}^{t}\left|\zeta(\tau)\right|^2 d\tau \le \int\limits_{t_0}^{t}\chi^T(\tau){\Upsilon}\chi(\tau) d\tau+\beta(t_0)
\end{equation}
holds along trajectories of $\Sigma_{12}^{\mathcal D}$, where $\beta(t_0)\ge 0$ may depend on particular values of the variables (manifest and latent) of $\Sigma_{12}^{\mathcal D}$ at time instant $t_0\in {\mathbb R}_+$. $\square$ 

In the above problem formulation, part i) implies that $\Sigma_{12}^{\mathcal D}$ possesses the same essential properties as $\Sigma_1$ and $\Sigma_2$ do, and therefore can be interconnected with other dissipative networks using similar methods. Part ii), on the other hand, implies the existence of an ${\mathcal L}_2$-gain from input variables $\chi$ to output variables $\zeta$. As $\zeta$ comprise all interconnection variables of $\Sigma_{12}^{\mathcal D}$, this in particular means that the interconnection variables within the network $\Sigma_{12}^{\mathcal D}$ remain bounded in the quadratic integral sense as long as the input variables are bounded.

%\newpage 

\section{Design of interconnection controller ${\mathcal D}_s$}\label{SectionDesignInterconnectionController01} 
\begin{figure}[h!]
    \begin{center}
        \includegraphics[scale=0.4]{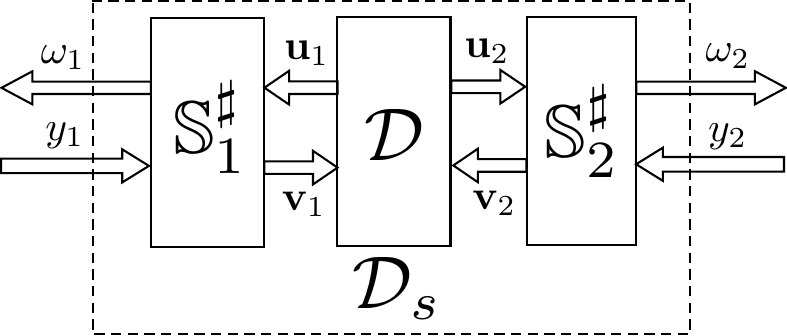}
       \caption{Scattering-based dynamic controller ${\mathcal D}_s$}
        \label{Fig:Figure05}
    \end{center}
\end{figure}

In this section, a detailed procedure for design of interconnection controller ${\mathcal D}_s$ is presented together with a number of technical results that justify the procedure. 
The structure of the interconnection controller ${\mathcal D}_s$ is shown in Figure~\ref{Fig:Figure05}. The controller comprises two scattering transformations ${\mathbb S}^\sharp_1$, ${\mathbb S}_2^\sharp$, as well as a dynamic block denoted by ${\mathcal D}$. 
For each $i=1,2$, the scattering transformation ${\mathbb S}^\sharp_i$ relate signals ${\bf u}_i$, $y_i$ to variables $\omega_i$, ${\bf v}_i$, according to the formula
\begin{equation}
\begin{bmatrix}
\omega_i\\ {\bf v}_i
\end{bmatrix}={\mathbb S}^\sharp_i
\begin{bmatrix}
{\bf u}_i\\  y_i
\end{bmatrix}, \hspace{0.5cm} i=1,2. \label{ScatteringVariablesSharpForS1S2}
\end{equation}
The dynamic controller ${\mathcal D}$ is an LTI system of the form
\begin{equation}
\begin{aligned}
{\dot x} & =  Ax+B {\bf v}\\
y & = Cx+D{\bf v}
\end{aligned}\label{ControllerD01}
\end{equation}
where 
\begin{equation}
{\bf v}:=\left[{\bf v}^T_1\, {\bf v}^T_2\right]^T\in{\mathbb R}^{\mathfrak p},\hspace{0.5cm} {\mathfrak p}:=p_1+p_2\label{Boldv}
\end{equation}
is the aggregated input to the controller~\eqref{ControllerD01},  $x\in{\mathbb R}^{\mathfrak n}$ is the controller state, where ${\mathfrak n}:=max\left\{ {\mathfrak m}, {\mathfrak p}\right\}$.
%, and ${\mathfrak m}:=m_1+m_2$. 
Also,   $y\in{\mathbb R}^{\mathfrak n}$ is the controller output, and matrices $A\in{\mathbb R}^{{\mathfrak n}\times {\mathfrak n}}$, $B\in{\mathbb R}_+^{{\mathfrak n}\times {\mathfrak p}}$, $C\in{\mathbb R}_+^{{\mathfrak n}\times {\mathfrak n}}$ and $D\in{\mathbb R}_+^{{\mathfrak n}\times {\mathfrak p}}$ are to be designed. The output $y$ of the controller~\eqref{ControllerD01} is connected to scattering transformations ${\mathbb S}_1$ and ${\mathbb S}_2$ according to the formula
\begin{equation}
{\bf u}  =  {\mathcal W} y, 
\label{WeightingMatrixW01}
\end{equation}
where 
\begin{equation}
{\bf u}:=\left[{\bf u}^T_1\, {\bf u}^T_2\right]^T\in{\mathbb R}^{\mathfrak m}, \hspace{0.5cm} \label{Boldu}
\end{equation} 
is the aggregated input of  ${\mathbb S}_1$ and ${\mathbb S}_2$, and ${\mathcal W}\in{\mathbb R}_+^{{\mathfrak m}\times {\mathfrak n}}$ is a weighting matrix with nonnegative entries to be designed. The design process for ${\mathbb S}_1$, ${\mathbb S}_2$, ${\mathcal D}$, and ${\mathcal W}$ is described below.

\subsection{Design of ${\mathbb S}_1^\sharp$, ${\mathbb S}_2^\sharp$}In order to design transformations ${\mathbb S}_1^\sharp$, ${\mathbb S}_2^\sharp$ in~\eqref{ScatteringVariablesSharpForS1S2}, we begin with the design of scattering transformations ${\mathbb S}_1$, ${\mathbb S}_2$ of a more conventional form
\begin{equation}
\begin{bmatrix}
\omega_i\\ y_i
\end{bmatrix}={\mathbb S}_i
\begin{bmatrix}
{\bf u}_i\\ {\bf v}_i
\end{bmatrix}, \hspace{0.5cm} i=1,2. \label{ScatteringVariablesForS1S2}
\end{equation}
which relate signals $\omega_i$, $y_i$ to the scattering variables ${\bf u}_i$, ${\bf v}_i$. Subsequently, ${\mathbb S}_1^\sharp$, ${\mathbb S}_2^\sharp$ will be designed by resolving~\eqref{ScatteringVariablesForS1S2} in the form~\eqref{ScatteringVariablesSharpForS1S2}, where the latter calculates the output variables $\omega_i$,  ${\bf v}_i$ as a (linear) function of the input variables  ${\bf u}_i$, $y_i$. 

 By Assumption~\ref{AssumptionDissipativity01}, networks $\Sigma_1$, $\Sigma_2$ are dissipative with quadratic supply rates with matrices $W_1=W_1^T\in{\mathbb R}^{{\mathcal N}_1\times {\mathcal N}_1}$, $W_2=W_2^T\in{\mathbb R}^{{\mathcal N}_2\times {\mathcal N}_2}$. Let $\Lambda_i\in{\mathbb R}^{{\mathcal N}_i\times {\mathcal N}_i}$, $i=1,2$, be diagonal matrices whose main diagonals comprise eigenvalues of $W_i$, $i=1,2,$, respectively, written in the descending order. Furthermore, let ${G}_i\in{\mathbb R}^{{\mathcal N}_i\times {\mathcal N}_i}$, $i=1,2$, be matrices whose columns are orthonormal eigenvectors of $W_i$, $i=1,2,$ respectively, such that ${G}_i\Lambda_i= W_i{G}_i$, $i=1,2$. Matrices of the scattering transformations ${\mathbb S}_1$, ${\mathbb S}_2$ can then be chosen as follows
 \begin{equation}\label{ScatteringTransS1S2}
 {\mathbb S}_i={G}_i P_i, \hspace{0.5cm} i=1,2,
 \end{equation}
 where 
 \begin{equation}\label{Pi01}
P_i:=\begin{bmatrix}
P^{+}_i & {\mathbb O}\\
{\mathbb O} & P^{-}_i
\end{bmatrix}, \hspace{0.5cm} i=1,2,
\end{equation}
and $P^{+}_i\in{\mathbb R}^{m_i\times m_i}$, $P^{-}_i\in{\mathbb R}^{p_i\times p_i}$, $i=1,2,$ are arbitrary permutation matrices. The use of permutation matrices in~\eqref{ScatteringTransS1S2} allows for additional flexibility in the design of scattering transformations ${\mathbb S}_1$, ${\mathbb S}_2$.

In order to resolve transformations~\eqref{ScatteringTransS1S2} in the form~\eqref{ScatteringVariablesSharpForS1S2}, let us first note that combining~\eqref{ScatteringTransS1S2}, \eqref{Pi01}, one can write 
\begin{equation}
 {\mathbb S}_i=
\begin{bmatrix}
G_{11}^{\{i\}}P^{+}_i & G_{12}^{\{i\}}P^{-}_i\\
G_{21}^{\{i\}}P^{+}_i & G_{22}^{\{i\}}P^{-}_i
\end{bmatrix}, \hspace{0.5cm} i=1,2. \label{ScatteringVariablesForS1S2BlockDiagonal}
\end{equation}
By Assumption~\ref{AssumptionRealizability01}, $G_{22}^{\{i\}}$ are full rank, and therefore~\eqref{ScatteringVariablesForS1S2} is equivalent to~\eqref{ScatteringVariablesSharpForS1S2} with 
\begin{equation}
{\mathbb S}_i^{\sharp}
=\begin{bmatrix}
\left(G_{11}^{\{i\}}- G_{12}^{\{i\}}\left(G_{22}^{\{i\}}\right)^{-1}G_{21}^{\{i\}} \right)P^{+}_i 
& G_{12}^{\{i\}} \left(G_{22}^{\{i\}}\right)^{-1}\\
- \left(G_{22}^{\{i\}}P^{-}_i\right)^{-1}G_{21}^{\{i\}}P^{+}_i & \left(G_{22}^{\{i\}}P^{-}_i\right)^{-1}
\end{bmatrix}, \hspace{0.5cm} i=1,2. \label{S1S2Sharp}
\end{equation}
For more details, see Appendix~\ref{AppendixRealizability01}. Now, note that Assumption~\ref{AssumptionLiveness01} together with the structure of the permutation matrices $P_1$, $P_2$ imply that matrices $P_i^T\Lambda_iP_i$ are of the form
\begin{equation}\label{Lambda+Lambda-a}
P_i^T\Lambda_iP_i:=\begin{bmatrix}
\Lambda^{+}_i & {\mathbb O}\\
{\mathbb O} & -\Lambda^{-}_i
\end{bmatrix}, \hspace{0.5cm} i=1,2,
\end{equation}
where $\Lambda^{+}_i\in {\mathbb R}^{m_i\times m_i}$, $\Lambda^{-}_i\in {\mathbb R}^{p_i\times p_i}$ are diagonal matrices such that $\Lambda^{+}_i\succeq 0$, $\Lambda^{-}_i\succ 0$, $i=1,2$. The main diagonals of  $\Lambda^{+}_i$ and $-\Lambda^{-}_i$ comprise the nonnegative and negative eigenvalues, respectively, of $W_i$, $i=1,2$.  
In the following, we will use notation:
\begin{equation}\label{LambdaPlusMinus01}
\Lambda^+:=\begin{bmatrix}
\Lambda^{+}_1 & {\mathbb O}\\
{\mathbb O} & \Lambda^{+}_2
\end{bmatrix}\in {\mathbb R}^{{\mathfrak m}\times {\mathfrak m}}
, \hspace{0.5cm} \Lambda^-:=\begin{bmatrix}
\Lambda^{-}_1 & {\mathbb O}\\
{\mathbb O} & \Lambda^{-}_2
\end{bmatrix}\in {\mathbb R}^{{\mathfrak p}\times {\mathfrak p}}.
\end{equation}

\subsection{Design of ${\mathcal D}$: Technical Results} %Denote ${\mathfrak m}=m_1+m_2$, ${\mathfrak p}=p_1+p_2$, and ${\mathfrak n}=max\left\{ {\mathfrak m}, {\mathfrak p}\right\}$. 
Below some technical results will be presented which, in particular, will be used for justification of the design procedure for interconnection controller ${\mathcal D}$. 
First, pick an arbitrary $A_0\in{\mathbb R}^{{\mathfrak n}\times {\mathfrak n}}$ that satisfies the Lyapunov equation
\begin{equation}\label{LyuapunovEquation01}
A_0^T {\mathcal P}+{\mathcal P}A_0=-{\mathcal Q}
\end{equation}
for some ${\mathcal Q}={\mathcal Q}^T\succ 0$ and ${\mathcal P}={\mathcal P}^T\succ 0$. Also, pick an arbitrary full rank $B_0\in{\mathbb R}^{{\mathfrak n}\times {\mathfrak p}}$, and form a matrix
\begin{equation}\label{W0}
W_0=
\begin{bmatrix}
{\mathbb O} & B_0^T{\mathcal P}\\
{\mathcal P}B_0 & -{\mathcal Q}
\end{bmatrix}
\end{equation}

\lemma\label{LemmaA0B0} Matrix $W_0=W_0^T$ has exactly ${\mathfrak n}$ negative ($<0$) eigenvalues and ${\mathfrak p}$ positive ($>0$) eigenvalues. $\square$

Proof of Lemma~\ref{LemmaA0B0} is given in Appendix~\ref{AppendixProofs01}. Next, let $G\in{\mathbb R}^{({\mathfrak n}+{\mathfrak p})\times ({\mathfrak n}+{\mathfrak p})}$ be an orthonormal matrix such that 
\begin{equation}\label{LambdaD01}
W_0G=G\Lambda_d,
\end{equation}
where $\Lambda_d:=\mbox{ diag }\left\{\lambda^d_1 , \ldots , \lambda^d_{{\mathfrak n}+{\mathfrak p}} \right\}$ is a diagonal matrix whose main diagonal comprises the eigenvalues of $W_0$ written in the descending order, $\lambda^d_1\ge  \ldots \ge \lambda^d_{\mathfrak p} > 0 > \lambda^d_{{\mathfrak p}+1}\ge \ldots \ge  \lambda^d_{{\mathfrak n}+{\mathfrak p}}$. In the following, we will use notation $\Lambda_d^+:=\mbox{ diag }\left\{\lambda^d_1 , \ldots , \lambda^d_{\mathfrak p} \right\}\succ 0$, and $\Lambda_d^-:=\mbox{ diag }\left\{-\lambda^d_{{\mathfrak p}+1} , \ldots , -\lambda^d_{{\mathfrak p}+{\mathfrak n}} \right\}\succ 0$; thus, %$\Lambda^D=\left\{\Lambda^D_+, -\Lambda^D_-\right\}$. 
\begin{equation*}%\label{LambdaDDiagonal20}
\Lambda_d=\begin{bmatrix}
\Lambda_d^+ & {\mathbb O}\\
{\mathbb O} &-\Lambda_d^- 
\end{bmatrix}. 
\end{equation*}
Consider a partition of $G$ of the form
\begin{equation}\label{PartitionG01}
G=
\begin{bmatrix}
{G}_{11} &  {G}_{12} \\
{G}_{21} &  {G}_{22} 
\end{bmatrix},
\end{equation}
where ${G}_{11}\in{\mathbb R}^{{\mathfrak p}\times {\mathfrak p}}$, ${G}_{22}\in{\mathbb R}^{{\mathfrak n}\times {\mathfrak n}}$, and ${G}_{21}, {G}^T_{12}\in{\mathbb R}^{{\mathfrak n}\times {\mathfrak p}}$. The following result is valid. 

\lemma\label{LemmaRankG22} Rank ${G}_{22}={\mathfrak n}$. $\square$

Proof of Lemma~\ref{LemmaRankG22} can be found in Appendix~\ref{ProofLemmaG22Rank}. 
Now,  define
\begin{equation}\label{LambdaStar01}
\Lambda_*:= 
\begin{bmatrix}
\Lambda_*^{+} & {\mathbb O}\\
{\mathbb O} & -\Lambda_*^{-}
\end{bmatrix},
\end{equation}
where 
$\Lambda_*^{+} \in{\mathbb R}^{{\mathfrak p}\times {\mathfrak p}}$, $\Lambda_*^{-} \in{R}^{{\mathfrak n}\times {\mathfrak n}}$ are given diagonal positive definite matrices.
Next, define
\begin{equation}\label{PD01}
 {\rm P}_d:=\begin{bmatrix} P_d^{+} & {\mathbb O}\\
{\mathbb O} & P_d^{-} 
\end{bmatrix}
\end{equation} 
where $P_d^{+} \in{\mathbb R}^{{\mathfrak p}\times {\mathfrak p}}$, $P_d^{-} \in{\mathbb R}^{{\mathfrak n}\times {\mathfrak n}}$ are arbitrary permutation matrices. 
Finally, let $\Gamma_d\in{\mathbb R}^{({\mathfrak n}+{\mathfrak p})\times ({\mathfrak n}+{\mathfrak p})}$ be a diagonal positive definite matrix that satisfies 
\begin{equation}\label{GammaD01}
\Gamma^2_d:={\rm P}_d^T\Lambda_d^{-1} {\rm P}_d \Lambda_*. \end{equation}
The above definition of  $\Gamma_d$ is valid 
%where the right hand side is well defined 
because the expression in the right-hand side of~\eqref{GammaD01} is a positive definite diagonal matrix. 
Now, define 
\begin{equation}\label{GCal01}
{\mathcal G}:= G\, {\rm P}_d\Gamma_d\in{\mathbb R}^{({\mathfrak n}+{\mathfrak p})\times ({\mathfrak n}+{\mathfrak p})}, 
\end{equation} 
and consider the following partition of ${\mathcal G}$:
\begin{equation}\label{GCalPartitioned01}
{\mathcal G}=
\begin{bmatrix}
{\mathcal G}_{11} &  {\mathcal G}_{12} \\
{\mathcal G}_{21} &  {\mathcal G}_{22} 
\end{bmatrix},
\end{equation}
where 
${\mathcal G}_{11}\in{\mathbb R}^{{\mathfrak p}\times {\mathfrak p}}$, ${\mathcal G}_{22}\in{\mathbb R}^{{\mathfrak n}\times {\mathfrak n}}$, and ${\mathcal G}_{21}, {\mathcal G}^T_{12}\in{\mathbb R}^{{\mathfrak n}\times {\mathfrak p}}$. Straightforward calculation gives  ${\mathcal G}_{22}={G}_{22}P_d^{-}\Gamma_d^-$, where $\Gamma_d^-$ is the bottom right ${\mathfrak n}\times {\mathfrak n}$ block of $\Gamma_d$; in particular, ${\mathcal G}_{22}$ is full rank as a product of square full-rank matrices. 
Now,  consider a controller of the form~\eqref{ControllerD01}, where
\begin{equation}\label{AClosedLoop01}
A:=A_0+B_0{\mathcal G}_{12} {\mathcal G}_{22}^{-1} 
\end{equation}
\begin{equation}\label{BClosedLoop01}
B:=B_0 \left( {\mathcal G}_{11} -{\mathcal G}_{12}{\mathcal G}_{22}^{-1} {\mathcal G}_{21}  \right)
\end{equation}
\begin{equation}\label{СDClosedLoop01}
C:={\mathcal G}_{22}^{-1}, \hspace{0.3cm} D=- {\mathcal G}_{22}^{-1} {\mathcal G}_{21} 
\end{equation}
The following result is valid. 

\lemma\label{LemmaDissipativityControllerWithY01} Consider a controller of the form~\eqref{ControllerD01}, where the controller's matrices are chosen according to the formulas~\eqref{AClosedLoop01}-\eqref{СDClosedLoop01}. Then:

i) The controller~\eqref{ControllerD01} is dissipative with supply rate
\begin{equation}\label{SupplyRateController05a}
s_d:=\begin{bmatrix}
{\bf v}\\ y
\end{bmatrix}^T
\Lambda_*
\begin{bmatrix}
{\bf v}\\ y
\end{bmatrix},
\end{equation}
where $\Lambda_*$ is defined by~\eqref{LambdaStar01}. 

ii) Matrix $A$ defined by~\eqref{AClosedLoop01} is stable.

{\bf Proof:} Consider a system 
\begin{equation}\label{SystemA0B0}
{\dot x}=A_0x+B_0\eta 
\end{equation}
where $A_0$ and $B_0$ are defined above. Taking into account~\eqref{LyuapunovEquation01}, it is straightforward to see that the time derivative of a positive definite storage function candidate 
\begin{equation}\label{DefinitionVD01}
V_d:=x^T{\mathcal P} x
\end{equation}
satisfies 
\begin{equation}\label{DotV01a}
{\dot V}_d=\begin{bmatrix}
\eta \\ x
\end{bmatrix}^T
W_0
\begin{bmatrix}
\eta \\ x
\end{bmatrix},
\end{equation}
where $W_0$ is defined by~\eqref{W0}. Now, consider a change of coordinates defined by the formula
\begin{equation}
\begin{bmatrix}
\eta \\ x
\end{bmatrix}
={\mathcal G}
\begin{bmatrix}
{\bf v} \\ y
\end{bmatrix}:=
\begin{bmatrix}
{\mathcal G}_{11} &  {\mathcal G}_{12} \\
{\mathcal G}_{21} &  {\mathcal G}_{22} 
\end{bmatrix}
\begin{bmatrix}
{\bf v} \\ y
\end{bmatrix}. \label{ScatteringTransform01aaa}
\end{equation}
We know that ${\mathcal G}_{22}$ is full rank, therefore the above transformation~\eqref{ScatteringTransform01aaa} implies that 
\begin{align}
\eta&=\left( {\mathcal G}_{11} -{\mathcal G}_{12}{\mathcal G}_{22}^{-1} {\mathcal G}_{21}  \right){\bf v}+{\mathcal G}_{12}{\mathcal G}_{22}^{-1} x\label{eta01}\\
y &={\mathcal G}_{22}^{-1} x - {\mathcal G}_{22}^{-1} {\mathcal G}_{21} {\bf v} \label{y01}
\end{align}
and combining~\eqref{SystemA0B0} with~\eqref{eta01}, \eqref{y01}, one obtains an LTI system of the form~\eqref{ControllerD01}, where matrices $A$, $B$, $C$, and $D$ are given by~\eqref{AClosedLoop01}-\eqref{СDClosedLoop01}. Thus, the system~\eqref{ControllerD01} with matrices $A$, $B$, $C$, $D$  given by~\eqref{AClosedLoop01}-\eqref{СDClosedLoop01} can be obtained as a result of input-output transformation~\eqref{ScatteringTransform01aaa} applied to system~\eqref{SystemA0B0}. Now, combining~\eqref{DotV01a} and~\eqref{ScatteringTransform01aaa}, one obtains
\begin{equation}\label{DotV02aa}
{\dot V}_d=
\begin{bmatrix}
{\bf v} \\ y
\end{bmatrix}^T{\mathcal G}^T
W_0
{\mathcal G}
\begin{bmatrix}
{\bf v} \\ y
\end{bmatrix}. 
\end{equation}
Taking into account~\eqref{GCal01}, \eqref{LambdaD01}, and \eqref{GammaD01}, 
one finds
\begin{equation*}
{\mathcal G}^T
W_0
{\mathcal G}=
\Gamma_d {\rm P}_d^TG^T
W_0
G {\rm P}_d\Gamma_d=
\Gamma_d {\rm P}_d^T\Lambda_d {\rm P}_d\Gamma_d=\Lambda_*,
\end{equation*}
 which combined with~\eqref{DotV02aa} proves~\eqref{SupplyRateController05a}. 

To prove part ii), set ${\bf v} = 0$. From~\eqref{SupplyRateController05a}, \eqref{LambdaStar01}, it follows that 
\begin{equation*}
{\dot V}_d=-y^T\Lambda_*^{-}y, 
\end{equation*}
and using~\eqref{y01} with ${\bf v} = 0$, one obtains 
\begin{equation*}
{\dot V}_d=-x^T{\hat Q}x, 
\end{equation*}
where ${\hat{\mathcal Q}}:=\left({\mathcal G}_{22}^{-1}\right)^T\Lambda_*^{-}{\mathcal G}_{22}^{-1}$. 
Matrix ${\hat{\mathcal Q}}$ is positive definite as a congruency transformation of a positive definite $\Lambda_*^{\{-\}}$. Taking into account~\eqref{DefinitionVD01}, we see that matrix $A$ defined by~\eqref{AClosedLoop01} satisfies 
\begin{equation}
A^T {\mathcal P}+{\mathcal P}A=-{\hat{\mathcal Q}}
\end{equation}
where both ${\mathcal P}\succ 0$, ${\hat{\mathcal Q}}\succ 0$, which implies that $A$ is stable (see for example~\cite[Lemma 3.19, ii)]{zhou:doyle:glover:96}). $\square$

Now, pick an arbitrary positive definite diagonal matrix $\Lambda_{\bf u}\in {\mathbb R}^{{\mathfrak m}\times {\mathfrak m}}$, and let a matrix ${\mathcal W}\in{\mathbb R}_+^{{\mathfrak m}\times {\mathfrak n}}$ in~\eqref{WeightingMatrixW01} be such that 
\begin{equation}\label{ConditionsOnWCal01}
\Lambda_*^{-}\succeq {\mathcal W}^T\Lambda_{\bf u} {\mathcal W}
\end{equation}
holds. The following result is a simple consequence of Lemma~\ref{LemmaDissipativityControllerWithY01}

 %Pick an arbitrary matrix ${\mathcal W}\in{\mathbb R}_+^{{\mathfrak m}\times {\mathfrak n}}$ in~\eqref{WeightingMatrixW01} that satisfies 

\corollary\label{Corollary010a} Let matrix ${\mathcal W}\in{\mathbb R}_+^{{\mathfrak m}\times {\mathfrak n}}$ in~\eqref{WeightingMatrixW01} satisfies~\eqref{ConditionsOnWCal01}. Then, under the conditions of Lemma~\ref{LemmaDissipativityControllerWithY01}, the controller~\eqref{ControllerD01} with ``output'' ${\bf u}$ defined by~\eqref{WeightingMatrixW01} is dissipative with supply rate
\begin{equation}\label{SupplyRateController010a}
{\hat s}_d:=- {\bf u}^T\Lambda_{\bf u}{\bf u} +{\bf v}^T\Lambda_*^{+}{\bf v}. 
\end{equation}

{\bf Proof:} We have 
\begin{equation*}\label{SupplyRateController06a}
s_d:=\begin{bmatrix}
{\bf v}\\ y
\end{bmatrix}^T
\Lambda_*
\begin{bmatrix}
{\bf v}\\ y
\end{bmatrix}= - {y}^T\Lambda_*^{-} y +{\bf v}^T\Lambda_*^{+} {\bf v}.
\end{equation*}
Taking into account~\eqref{ConditionsOnWCal01}, we have ${y}^T\Lambda_*^{-} y\ge {y}^T{\mathcal W}^T\Lambda_{\bf u} {\mathcal W} y= {\bf u}^T\Lambda_{\bf u}{\bf u}$, and therefore
\begin{equation*}
s_d= - {y}^T\Lambda_*^{-} y +{\bf v}^T\Lambda_*^{+} {\bf v}\le - {\bf u}^T\Lambda_{\bf u}{\bf u} +{\bf v}^T\Lambda_*^{+} {\bf v}:={\hat s}_d. \hspace{0.5cm} \square
\end{equation*}
%$\square$

\subsection{Design of ${\mathcal D}$: Procedure}
Based on the preliminary developments presented above,  the following procedure is proposed which deals with design of matrices  $A$, $B$, $C$, $D$ of the controller~\eqref{ControllerD01} and the output weighting matrix ${\mathcal W}$ in~\eqref{WeightingMatrixW01}, as well as auxiliary parameters $\delta>0$ and $\theta>0$ which will be used in the subsequent analysis of interconnection $\Sigma_{12}^{\mathcal D}$. 

\procedure\label{ProcedureDesign01}

{\bf  Given:} matrices $\Lambda^+$, $\Lambda^-$  defined by~\eqref{LambdaPlusMinus01}. 

{\bf Do} the following steps: 

\begin{enumerate}

\item\label{Step5} Pick an arbitrary $\delta>0$. 

\item Pick an arbitrary $\theta>0$ and arbitrary positive definite diagonal $\Lambda_*^{+} \in{\mathbb R}^{{\mathfrak p}\times {\mathfrak p}}$ such that 
\begin{equation}\label{Ineq01a}
\Lambda_*^{+}+\delta\,{ \mathbb I}_{\mathfrak p}\preceq \theta\Lambda^{-}.
\end{equation} 

\item Pick a diagonal matrix $\Lambda_{\bf u}\in {\mathbb R}^{{\mathfrak m}\times {\mathfrak m}}$ such that 
\begin{equation}
\Lambda_{\bf u}\succeq\theta\Lambda^{+}+\delta\,{ \mathbb I}.  \label{Ineq02a}
\end{equation}

\item Pick a weighting matrix ${\mathcal W}\in{\mathbb R}_+^{{\mathfrak m}\times {\mathfrak n}}$ in~\eqref{WeightingMatrixW01} and a diagonal $\Lambda_*^{-} \in{R}^{{\mathfrak n}\times {\mathfrak n}}$ such that
\eqref{ConditionsOnWCal01} holds. In particular, one can choose $\Lambda_*^{-}:=\sigma_{max}\,{\mathbb I}_{\mathfrak n}$, where $\sigma_{max}$ is the maximum singular value of ${\mathcal W}^T\Lambda_{\bf u} {\mathcal W}$. Form a matrix $\Lambda_*$ according to the formula~\eqref{LambdaStar01}.

\item\label{Step1} Pick an arbitrary $A_0\in{\mathbb R}^{{\mathfrak n}\times {\mathfrak n}}$ that satisfies the Lyapunov equation~\eqref{LyuapunovEquation01}
for some ${\mathcal Q}={\mathcal Q}^T\succ 0$ and ${\mathcal P}={\mathcal P}^T\succ 0$. Pick an arbitrary full rank $B_0\in{\mathbb R}^{{\mathfrak n}\times {\mathfrak p}}$. 

%\item Pick a diagonal matrix $\Lambda_{\bf u}\in {\mathbb R}^{{\mathfrak m}\times {\mathfrak m}}$, $\Lambda_{\bf u} \succ 0$ such that 
%\begin{equation}
%\Lambda_{\bf u}=\theta\Lambda^{+}+\delta\,{ \mathbb I}, \label{Ineq02a}
%\end{equation}

\item Form $W_0$ according to~\eqref{W0}, and calculate matrices $\Lambda_d$ and $G$ according to the formula~\eqref{LambdaD01}. %, \eqref{LambdaDDiagonal20}, and \eqref{PartitionG01}. 

\item Pick an arbitrary %matrices $\Lambda_*$ and 
${\rm P}_d$ of the form~\eqref{PD01}.
%~\eqref{LambdaStar01} and~\eqref{PD01}.
%, respectively. 

\item Calculate partitioned matrix ${\mathcal G}$ according to the formulas~\eqref{GammaD01}, \eqref{GCal01}, and \eqref{GCalPartitioned01}. 

\item Calculate matrices $A$, $B$, $C$, and $D$ of the controller~\eqref{ControllerD01} using formulas~\eqref{AClosedLoop01}, \eqref{BClosedLoop01}, \eqref{СDClosedLoop01}.

\end{enumerate}

{\bf End of Procedure~\ref{ProcedureDesign01}}\\
%\end{enumerate}

\remark\label{RemarkPermutationMatrices} Expressions for scattering transformations~\eqref{ScatteringTransS1S2}, \eqref{GCal01} contain permutation matrices ${\rm P}_1$, ${\rm P}_2$, and ${\rm P}_d$.
%, respectively. 
These permutation matrices can be chosen arbitrary as long as they are of the form~\eqref{Pi01} and \eqref{PD01}, respectively. Introduction of such permutation matrices into the scattering-based controllers allows for additional flexibility in the control design process, which can be utilized for the purpose of achieving additional goals such as performance optimization for example. $\square$

Specific dissipativity properties of the interconnection $\Sigma_{12}^{\mathcal D}$, where various parameters were designed using the above Procedure~\ref{ProcedureDesign01}, will be studied in the next section.

%\newpage 

\section{Dissipativity Properties of $\Sigma_{12}^{\mathcal D}$}\label{SectionDissipativityPropertiesInterconnection01}
The next result shows that the interconnection $\Sigma_{12}^{\mathcal D}$ comprising subsystems $\Sigma_1$, $\Sigma_2$ and the interconnection controller ${\mathcal D}_s$ possesses dissipativity properties with a well defined supply rate function.

\theorem\label{TheoremDissipativitySigma12D} Consider the interconnection $\Sigma_{12}^{\mathcal D}$, where subsystems $\Sigma_1$, $\Sigma_2$ satisfy Assumptions~\ref{AssumptionDissipativity01}, \ref{AssumptionLiveness01}, and~\ref{AssumptionRealizability01}. Suppose also that matrices of scattering transformations ${\mathbb S}_1$, ${\mathbb S}_2$ are designed according to~\eqref{ScatteringTransS1S2}. Suppose matrices $A$, $B$, $C$, and $D$ of the controller~\eqref{ControllerD01} as well as parameters  $\delta>0$, $\theta>0$  are chosen according to Procedure~\ref{ProcedureDesign01}. Then $\Sigma_{12}^{\mathcal D}$ is dissipative with supply rate 
\begin{equation}\label{s12Definition}
s_{12}:=\begin{bmatrix}
\chi \\ \zeta
\end{bmatrix}^T 
W_{12}
\begin{bmatrix}
\chi \\ \zeta
\end{bmatrix}
\end{equation} 
where
\begin{equation}\label{W12Definition}
W_{12}:=\begin{bmatrix}
\theta W_{\chi} & \theta W_{cross} \\
\theta W^T_{cross} & -\delta{ \mathbb I}_{{\mathcal N}_1+{\mathcal N}_2}
\end{bmatrix}\in{\mathbb R}^{{\mathfrak N}\times {\mathfrak N}}
\end{equation}
where the components of $W_{12}$ are 
\begin{equation*}
\begin{gathered}
W_{cross}:=
\begin{bmatrix}
{\tilde W}_{1} & {\tilde W}_{2}
\end{bmatrix}\in{\mathbb R}^{\mathfrak m\times({\mathcal N}_1+{\mathcal N}_2)}, \\
%\hspace{0.5cm}
{\tilde W}_{1}:=\begin{bmatrix}
{\mathbb I}_{m_1} & {\mathbb O}_{m_1\times p_1} \\
{\mathbb O}_{m_2\times m_1} & {\mathbb O}_{m_2\times p_1}
\end{bmatrix}
W_1\in{\mathbb R}^{{\mathfrak m}\times {\mathcal N}_1}, \\
%\hspace{0.5cm}
{\tilde W}_{2}:=\begin{bmatrix}
{\mathbb O}_{m_1\times m_2} & {\mathbb O}_{m_1\times p_2}\\
{\mathbb I}_{m_2} & {\mathbb O}_{m_2\times p_2}
\end{bmatrix}
W_2\in{\mathbb R}^{{\mathfrak m}\times{\mathcal N}_2}, 
\end{gathered}
\end{equation*}
and 
\begin{equation*}
W_{\chi}:=\begin{bmatrix}
{\mathbb I}_{m_1} & {\mathbb O}_{m_1\times p_1}\\
{\mathbb O}_{m_2\times m_1} & {\mathbb O}_{m_2\times p_1}
\end{bmatrix}
 W_1
 \begin{bmatrix}
{\mathbb I}_{m_1} & {\mathbb O}_{m_1\times m_2}\\
{\mathbb O}_{p_1\times m_1} & {\mathbb O}_{p_1\times m_2} 
\end{bmatrix}+
\begin{bmatrix}
{\mathbb O}_{m_1\times m_2}  & {\mathbb O}_{m_1\times p_2}\\
{\mathbb I}_{m_2} & {\mathbb O}_{m_2\times p_2}
\end{bmatrix}
W_2
\begin{bmatrix}
{\mathbb O}_{m_2\times m_1} & {\mathbb I}_{m_2}\\
{\mathbb O}_{p_2\times m_1} & {\mathbb O}_{p_2\times m_2} 
\end{bmatrix}\in{\mathbb R}^{{\mathfrak m}\times{\mathfrak m}}\bullet
\end{equation*}

{\bf Proof:} According to Assumption~\ref{AssumptionDissipativity01}, 
%implies that the following dissipation inequalities 
%\begin{equation*}
%{\mathcal V}_{i}\left( t\right)-{\mathcal V}_{i}\left( t_0\right)\le \int\limits_{t_0}^{t}s_i(\tau) d\tau, \hspace{0.3cm} i=1,2, 
%\end{equation*}
%hold along the trajectories of 
% 
subsystems $\Sigma_1$, $\Sigma_2$ are dissipative with supply rates $s_1$, $s_2$, respectively, where $s_1$ and $s_2$ are defined by~\eqref{FormulaSupplyRates01Assumption1}. Taking into account interconnection constraints~\eqref{InterconnectionConstraints01a}, for each $i=1,2,$ direct calculations give 
\begin{equation*}
\begin{gathered}
s_i:=\begin{bmatrix}
\eta_i\\ y_i
\end{bmatrix}^T W_i
\begin{bmatrix}
\eta_i\\ y_i
\end{bmatrix}
=
\begin{bmatrix}
\omega_i+\chi_i\\ y_i
\end{bmatrix}^T W_i
\begin{bmatrix}
\omega_i+\chi_i\\ y_i
\end{bmatrix}\\
=\begin{bmatrix}
\omega_i\\ y_i
\end{bmatrix}^T W_i
\begin{bmatrix}
\omega_i\\ y_i
\end{bmatrix}
+
2\begin{bmatrix}
\chi_i\\ 0
\end{bmatrix}^T W_i
\begin{bmatrix}
\omega_i\\ y_i
\end{bmatrix}
+
\begin{bmatrix}
\chi_i\\ 0
\end{bmatrix}^T W_i
\begin{bmatrix}
\chi_i\\ 0
\end{bmatrix}.  %\hspace{0.5cm} i=1,2,
\end{gathered}
\end{equation*}
Using change of variables~\eqref{ScatteringVariablesForS1S2}, where ${\mathbb S}_i$ are chosen according to~\eqref{ScatteringTransS1S2}, expressions for $s_i$, $i=1,2$ can be written in the form
\begin{equation*}
\begin{gathered}
s_i=\begin{bmatrix}
{\bf u}_i\\ {\bf v}_i
\end{bmatrix}^T \begin{bmatrix}
\Lambda^{+}_i & {\mathbb O}\\
{\mathbb O} & -\Lambda^{-}_i
\end{bmatrix}
\begin{bmatrix}
{\bf u}_i\\ {\bf v}_i
\end{bmatrix}
+
2\begin{bmatrix}
\chi_i\\ 0
\end{bmatrix}^T W_i
\begin{bmatrix}
\omega_i\\ y_i
\end{bmatrix}
+
\begin{bmatrix}
\chi_i\\ 0
\end{bmatrix}^T W_i
\begin{bmatrix}
\chi_i\\ 0
\end{bmatrix},
\end{gathered}
\end{equation*}
where we use notation introduced in~\eqref{Lambda+Lambda-a}. Next, 
\begin{equation*}
\begin{gathered}
s_1+s_2= \sum_{i=1}^2 \begin{bmatrix}
{\bf u}_i\\ {\bf v}_i
\end{bmatrix}^T \begin{bmatrix}
\Lambda^{+}_i & {\mathbb O}\\
{\mathbb O} & -\Lambda^{-}_i
\end{bmatrix}
\begin{bmatrix}
{\bf u}_i\\ {\bf v}_i
\end{bmatrix}
+
2\begin{bmatrix}
\chi_i\\ 0
\end{bmatrix}^T W_i
\begin{bmatrix}
\omega_i\\ y_i
\end{bmatrix}
+
\begin{bmatrix}
\chi_i\\ 0
\end{bmatrix}^T W_i
\begin{bmatrix}
\chi_i\\ 0
\end{bmatrix}\\
= 
\begin{bmatrix}
{\bf u}_1\\ {\bf u}_2\\ {\bf v}_1\\ {\bf v}_2
\end{bmatrix}^T \begin{bmatrix}
\Lambda^{+}_1 & {\mathbb O} & {\mathbb O} & {\mathbb O}\\
{\mathbb O} & \Lambda^{+}_2 & {\mathbb O} & {\mathbb O}\\
{\mathbb O} & {\mathbb O} & -\Lambda^{-}_1 & {\mathbb O}\\
 {\mathbb O} & {\mathbb O} & {\mathbb O} & -\Lambda^{-}_2 
\end{bmatrix}
\begin{bmatrix}
{\bf u}_1\\ {\bf u}_2\\ {\bf v}_1\\ {\bf v}_2
\end{bmatrix} 
+
2\begin{bmatrix}
\chi_1\\ \chi_2
\end{bmatrix}^T 
\begin{bmatrix}
{\mathbb I} & {\mathbb O}\\
{\mathbb O} & {\mathbb O}
\end{bmatrix}
W_1\begin{bmatrix}
\omega_1\\ y_1
\end{bmatrix}
+
2\begin{bmatrix}
\chi_1\\ \chi_2
\end{bmatrix}^T 
\begin{bmatrix}
{\mathbb O} & {\mathbb O}\\
{\mathbb I} & {\mathbb O}
\end{bmatrix}
W_2\begin{bmatrix}
\omega_2\\ y_2
\end{bmatrix}\\
+
\begin{bmatrix}
\chi_1\\ \chi_2
\end{bmatrix}^T 
\begin{bmatrix}
{\mathbb I} & {\mathbb O}\\
{\mathbb O} & {\mathbb O}
\end{bmatrix}
 W_1
 \begin{bmatrix}
{\mathbb I} & {\mathbb O}\\
{\mathbb O} & {\mathbb O}
\end{bmatrix}
\begin{bmatrix}
\chi_1\\ \chi_2
\end{bmatrix}^T 
+
\begin{bmatrix}
\chi_1\\ \chi_2
\end{bmatrix}^T 
\begin{bmatrix}
{\mathbb O} & {\mathbb O}\\
{\mathbb I} & {\mathbb O}
\end{bmatrix}
W_2
\begin{bmatrix}
{\mathbb O} & {\mathbb I}\\
{\mathbb O} & {\mathbb O}
\end{bmatrix}
\begin{bmatrix}
\chi_1\\ \chi_2
\end{bmatrix}\\
\end{gathered}
\end{equation*}
 \begin{equation*}
 \begin{gathered}
= 
\begin{bmatrix}
{\bf u} \\ {\bf v}
\end{bmatrix}^T \begin{bmatrix}
\Lambda^{+} & {\mathbb O}\\
{\mathbb O} &  -\Lambda^{-} 
\end{bmatrix}
\begin{bmatrix}
{\bf u} \\ {\bf v}
\end{bmatrix} 
+2\chi^T W_{cross} \zeta+\chi^T W_{\chi} \chi
 \end{gathered}
 \end{equation*}
 where ${\bf u}$, ${\bf v}$, $\chi$, and $\zeta$ are defined by~\eqref{Boldu}, \eqref{Boldv}, \eqref{Chi12}, and~\eqref{Zeta12}, respectively. 

Now, consider a storage function candidate of the form
\begin{equation}\label{OverallStoragFunction01}
{\mathcal V}:=\theta\cdot\left( {\mathcal V}_1+{\mathcal V}_2\right)+{V}_d, 
\end{equation}
where $V_1$ and $V_2$ are storage functions of dissipative systems $\Sigma_1$ and $\Sigma_2$, respectively, $V_D$ is defined by~\eqref{DefinitionVD01}, and  $\theta>0$ is the parameter chosen in Step~\ref{Step5} of Procedure~\ref{ProcedureDesign01}.  One has 
\begin{equation}\label{DissipationInequalityOverallStoragFunction01}
{\mathcal V}\left( t\right)-{\mathcal V}\left( t_0\right)\le \int_{t_0}^{t}\left(\theta\cdot \left( s_1+s_2\right)+{\hat s}_d\right) d\tau, 
\end{equation}
where 
\begin{equation*}
\theta\cdot \left( s_1+s_2\right)+{\hat s}_d
= 
\begin{bmatrix}
{\bf u} \\ {\bf v}
\end{bmatrix}^T \begin{bmatrix}
\theta\Lambda^{+} -\Lambda_{\bf u} & {\mathbb O}\\
{\mathbb O} &  \Lambda^*_{+}-\theta\Lambda^{-} 
\end{bmatrix}
\begin{bmatrix}
{\bf u} \\ {\bf v}
\end{bmatrix} 
+2\theta\chi^T W_{cross} \zeta+\theta\chi^T W_{\chi} \chi
\end{equation*}
Taking into account~\eqref{Ineq01a} and~\eqref{Ineq02a}, one concludes that $\Lambda^*_{+}-\theta\Lambda^{-}\preceq -\delta{\mathbb I}$ and $\theta\Lambda^{+} -\Lambda_{\bf u}\preceq -\delta{\mathbb I}$, respectively. Therefore
\begin{equation*}
\theta\cdot\left( s_1+s_2\right)+{\hat s}_d\le -\delta\begin{bmatrix}
{\bf u} \\ {\bf v}
\end{bmatrix}^T 
\begin{bmatrix}
{\bf u} \\ {\bf v}
\end{bmatrix} 
+2\theta\chi^T W_{cross} \zeta+\theta\chi^T W_{\chi} \chi
%\le 
%\begin{bmatrix}
%\chi \\ {\bf u} \\ {\bf v}
%\end{bmatrix}^T 
%\begin{bmatrix}
%W_{\chi} & \theta W_{cross} \\
%\theta W^T_{cross} & -2\,\delta{ \mathbb I}
%\end{bmatrix}
%\begin{bmatrix}
%\chi \\ {\bf u} \\ {\bf v}
%\end{bmatrix}
\end{equation*}
It is easy to verify that 
\begin{equation*}
\begin{bmatrix}
{\bf u} \\ {\bf v}
\end{bmatrix}^T 
\begin{bmatrix}
{\bf u} \\ {\bf v}
\end{bmatrix} =\zeta^T\zeta
\end{equation*}
Indeed,
\begin{equation*}
\begin{bmatrix}
{\bf u} \\ {\bf v}
\end{bmatrix}^T 
\begin{bmatrix}
{\bf u} \\ {\bf v}
\end{bmatrix} =
\begin{bmatrix}
{\bf u}_1\\ {\bf u}_2 \\ {\bf v}_1 \\ {\bf v}_2
\end{bmatrix}^T 
\begin{bmatrix}
{\bf u}_1\\ {\bf u}_2 \\ {\bf v}_1 \\ {\bf v}_2
\end{bmatrix} =
\begin{bmatrix}
{\bf u}_1 \\ {\bf v}_1 \\ {\bf u}_2 \\ {\bf v}_2
\end{bmatrix}^T 
\begin{bmatrix}
{\bf u}_1 \\ {\bf v}_1 \\ {\bf u}_2 \\ {\bf v}_2
\end{bmatrix}=
\begin{bmatrix}
\omega_1 \\ y_1 \\ \omega_2 \\ y_2
\end{bmatrix}^T
\begin{bmatrix}
{\mathbb S}_1 & {\mathbb O}\\
{\mathbb O} & {\mathbb S}_2
\end{bmatrix}
\begin{bmatrix}
{\mathbb S}_1^T & {\mathbb O}\\
{\mathbb O} & {\mathbb S}_2^T
\end{bmatrix}
\begin{bmatrix}
\omega_1 \\ y_1 \\ \omega_2 \\ y_2
\end{bmatrix}
=
\begin{bmatrix}
\omega_1 \\ y_1 \\ \omega_2 \\ y_2
\end{bmatrix}^T
\begin{bmatrix}
\omega_1 \\ y_1 \\ \omega_2 \\ y_2
\end{bmatrix}
=\zeta^T\zeta
\end{equation*}
Therefore, 
\begin{equation}
\begin{gathered}
\theta\left( s_1+s_2\right)+s_D\le -\delta\zeta^T\zeta
+2\theta\chi^T W_{cross} \zeta+\theta\chi^T W_{\chi} \chi
\\
= 
\begin{bmatrix}
\chi \\ \zeta
\end{bmatrix}^T 
W_{12}
\begin{bmatrix}
\chi \\ \zeta
\end{bmatrix}
%= 
%\begin{bmatrix}
%\chi_{12} \\ \zeta
%\end{bmatrix}^T 
%\begin{bmatrix}
%\theta W_{\chi} & \theta W_{cross} \\
%\theta W^T_{cross} & -\delta{ \mathbb I}
%\end{bmatrix}
%\begin{bmatrix}
%\chi_{12} \\ \zeta
%\end{bmatrix}
\end{gathered}
\label{QSREstimate01}
\end{equation}
The statement follows by combining~\eqref{DissipationInequalityOverallStoragFunction01} and~\eqref{QSREstimate01}. $\bullet$ 

\corollary\label{CorollaryDissipativitySigma12D} Under the conditions of Theorem~\ref{TheoremDissipativitySigma12D}, there exists 
${\hat W}_{12}={\hat W}_{12}^T\in{\mathbb R}^{{\mathfrak N}\times {\mathfrak N}}$ such that 

i) ${\hat W}_{12}$ is full rank, and $n_-( {\hat W}_{12} )= {\mathcal N}_1+{\mathcal N}_2$;

ii) Interconnection $\Sigma_{12}^{\mathcal D}$ is dissipative with storage function~\eqref{OverallStoragFunction01} and supply rate
\begin{equation}\label{SupplyRateHatS12}
{\hat s}_{12}:= 
\begin{bmatrix}
\chi \\ \zeta
\end{bmatrix}^T 
{\hat W}_{12}
\begin{bmatrix}
\chi \\ \zeta
\end{bmatrix}
\end{equation}

iii) ${\hat W}_{12}$ satisfies the realizability condition (Assumption~\ref{AssumptionRealizability01}). 

{\bf Proof:} First, denote $\lambda_*$ the minimum eigenvalue of a (symmetric) matrix $W_{\chi} + \theta\delta^{-1} W_{cross} W^T_{cross}$, and choose an arbitrary $\epsilon_0\ge 0$ such that $\epsilon_0+\lambda_*>0$. Define
\begin{equation*}
{\hat W}_{12}:=\begin{bmatrix}
\theta \left(W_{\chi}+\epsilon_0{\mathbb I}\right) & \theta W_{cross} \\
\theta W^T_{cross} & -\delta{ \mathbb I}
\end{bmatrix}.
\end{equation*}
First, it is clear that ${W}_{12}\preceq {\hat W}_{12}$, and therefore dissipativity with supply rate~\eqref{s12Definition} implies that with~\eqref{SupplyRateHatS12}. Second, 
calculating the Schur complement of ${\hat W}_{12}$ with respect to its bottom right block, one has 
\begin{equation}\label{WS12}
\begin{gathered}
{\hat W}^S_{12}=
\begin{bmatrix}
{\mathbb I} & \delta^{-1}\theta W_{cross} \\
{\mathbb O}  & { \mathbb I}
\end{bmatrix}
\begin{bmatrix}
\theta \left(W_{\chi}+\epsilon_0{\mathbb I}\right) & \theta W_{cross} \\
\theta W^T_{cross} & -\delta{ \mathbb I}
\end{bmatrix}
\begin{bmatrix}
{\mathbb I} & {\mathbb O}  \\
\delta^{-1}\theta W^T_{cross} & { \mathbb I}
\end{bmatrix}\\
=\begin{bmatrix}
\theta \left(W_{\chi}+\epsilon_0{\mathbb I}\right) + \theta^2\delta^{-1} W_{cross} W^T_{cross} & {\mathbb O} \\
{\mathbb O}  & -\delta{ \mathbb I}
\end{bmatrix}. 
\end{gathered} 
\end{equation}
Matrices ${\hat W}_{12}$ and ${\hat W}^S_{12}$ are congruent and, by Sylvester law of inertia~\cite[Theorem 4.5.8]{horn:johnson:13}, they have the same number of negative eigenvalues  and the same number of positive eigenvalues. Matrix ${\hat W}^S_{12}$ is block diagonal and comprises two nonzero blocks where, by construction, the top left block is positive definite and the bottom right block is negative definite. This means that ${\hat W}^S_{12}$ (and consequently ${\hat W}_{12}$) has ${\mathfrak m}$ positive ($>0$) eigenvalues and ${\mathcal N}_1+{\mathcal N}_2$ negative ($<0$) eigenvalues; in particular, ${\hat W}_{12}$ is full rank. Finally, to prove statement iii) of the Corollary, let ${\hat G}$ be an orthonormal matrix whose columns are eigenvectors of ${\hat W}_{12}$, specifically, ${\hat W}_{12}{\hat G}={\hat G}{\hat\Lambda}$, where ${\hat\Lambda}$ is a diagonal matrix with elements on the main diagonal equal to eigenvalues of ${\hat W}_{12}$ written in the descended order. The last equation immediately implies that ${\hat W}_{12}={\hat G}{\hat\Lambda}{\hat G}^T$. Now, direct calculation gives 
\begin{equation*}
\begin{bmatrix}
{\mathbb O}\\ {\mathbb I}_{{\mathcal N}_1+{\mathcal N}_2}
\end{bmatrix}^T
{\hat W}_{12}
\begin{bmatrix}
{\mathbb O}\\ {\mathbb I}_{{\mathcal N}_1+{\mathcal N}_2}
\end{bmatrix}= -\delta{ \mathbb I}
\end{equation*}
and therefore
\begin{equation*}
\begin{bmatrix}
{\mathbb O}\\ {\mathbb I}
\end{bmatrix}^T
\begin{bmatrix}
G_{11} & G_{12}\\
G_{21} & G_{22}
\end{bmatrix} 
\begin{bmatrix}
{\hat\Lambda}_+ & {\mathbb O}\\ 
{\mathbb O} & {\hat\Lambda}_-
\end{bmatrix}
\begin{bmatrix}
G^T_{11} & G^T_{21}\\
G_{12}^T & G^T_{22}
\end{bmatrix} 
\begin{bmatrix}
{\mathbb O}\\ {\mathbb I}
\end{bmatrix}
=
G_{21} {\hat\Lambda}_+G^T_{21}
+ G_{22}{\hat\Lambda}_-G^T_{22}
= -\delta{ \mathbb I}. 
\end{equation*}
Since $G_{21} {\hat\Lambda}_+G^T_{21}\succeq 0$, one sees that $G_{22}{\hat\Lambda}_-G^T_{22}\preceq -\delta{ \mathbb I}$, which implies that $G_{22}$ is full rank.  The proof is complete. $\bullet$

\corollary\label{CorollaryFiniteL2GainOfSigmaD12}Under the conditions of Theorem~\ref{TheoremDissipativitySigma12D}, there exist
${\Upsilon}_{12}={\Upsilon}_{12}^T\in {\mathbb R}^{{\mathfrak m}\times {\mathfrak m}}$, ${\Upsilon}_{12}\succeq 0$, and $\beta_0> 0$ such that the inequality 
\begin{equation}\label{InequalityL2GainOfSigmaD12}
\int\limits_{t_0}^{t}\left|\zeta(\tau)\right|^2 d\tau \le \int\limits_{t_0}^{t}\chi^T(\tau){\Upsilon}_{12}\chi(\tau) d\tau+\beta_0\cdot {\mathcal V}(t_0)
\end{equation}
holds for all $0\le t_0\le t_1$. 

{\bf Proof:} Theorem~\ref{TheoremDissipativitySigma12D} states that the dissipation inequality 
\begin{equation*}
{\mathcal V}\left( t\right)-{\mathcal V}\left( t_0\right)\le \int\limits_{t_0}^{t}s_{12}(\tau)  d\tau
\end{equation*}
holds for all $0\le t_0\le t$, where $s_{12}$ is defined by~\eqref{s12Definition}, \eqref{W12Definition}, and ${\mathcal V}$ by~\eqref{OverallStoragFunction01}. 
On the other hand, using Young's quadratic inequality $2 a^Tb\le a^T Q a+b^TQ^{-1}b$ which holds for any $Q=Q^T\succ 0$, one has
\begin{equation*}
2\theta\chi^T W_{cross} \zeta\le \frac{2\theta^2}{\delta} \chi^T W_{cross} W^T_{cross} \chi+\frac{\delta}{2} \zeta^T \zeta,
\end{equation*}
and therefore
\begin{equation*}
{\mathcal V}\left( t\right)-{\mathcal V}\left( t_0\right)\le 
\int\limits_{t_0}^t \begin{bmatrix}
\chi \\ \zeta
\end{bmatrix}^T 
\begin{bmatrix}
\theta W_{\chi} +\frac{2\theta^2}{\delta} W_{cross} W_{cross}^T & {\mathbb O} \\
{\mathbb O} & -\frac{\delta}{2}{ \mathbb I}
\end{bmatrix}
\begin{bmatrix}
\chi \\ \zeta
\end{bmatrix}
 d\tau.
\end{equation*}
Multiplying both sides of the last inequality by $2/\delta$ and taking into account that ${\mathcal V}\left( t\right)\ge 0$, one concludes that~\eqref{InequalityL2GainOfSigmaD12} holds with 
\begin{equation*}
{\Upsilon}_{12}:=\frac{2\theta}{\delta}W_{\chi} +\frac{4\theta^2}{\delta^2} W_{cross} W_{cross}^T 
\end{equation*}
and $\beta:=2/\delta$. The proof is complete. $\bullet$

Corollaries~\ref{CorollaryDissipativitySigma12D} and~\ref{CorollaryFiniteL2GainOfSigmaD12} established dissipativity properties of the interconnection 
$\Sigma_{12}^{\mathcal D}$ equipped with input $\chi:=\left[\chi_1\, \ldots \,  \chi_{{\mathfrak m}}\right]^T$ and output $\zeta:=\left[\zeta_1\, \ldots  \,  \zeta_{{\mathcal N}_1+{\mathcal N}_2}\right]^T$. As a final step, we will show that if, instead of the full input $\chi$ and full output $\zeta$,  only  (nonempty) subsets of input and output components are available as manifest variables, then the interconnection $\Sigma_{12}^{\mathcal D}$ with these new input and output of reduced dimensions remains dissipative with a quadratic supply rate, whose matrix satisfies all the same properties established in Corollaries~\ref{CorollaryDissipativitySigma12D} and~\ref{CorollaryFiniteL2GainOfSigmaD12} for the case of the full set of inputs and outputs. Specifically, consider the new input ${\bar\chi}:=\left[\chi_{i_1}\, \ldots ,\,  \chi_{i_{\bar m}}\right]^T$ and the new output ${\bar\zeta}:=\left[\zeta_{j_1}\, \ldots  \zeta_{j_{\bar n}}\right]^T$, where $1\le i_1<\ldots < i_{\bar m}\le {\mathfrak m}$, and $1\le j_1<\ldots < j_{\bar n}\le  {\mathcal N}_1+{\mathcal N}_2$. In words, ${\bar\chi}$ and ${\bar\zeta}$ represent subsets of input variables $\chi$ and output variables $\zeta$, respectively. Also, let ${\tilde\chi}$ be a vector that comprises components of $\chi$ that are not in ${\bar\chi}$; ${\tilde\zeta}$ is defined analogously. The following result is valid. 

\corollary\label{CorollaryDissipativitySigma12DReducedInputsOutputs01} Consider interconnection $\Sigma_{12}^{\mathcal D}$ where ${\tilde\chi}(\cdot)\equiv 0$. Under condition of Theorem~\ref{TheoremDissipativitySigma12D}, the system is dissipative with a supply rate 
\begin{equation}\label{SupplyRateBarS12}
{\bar s}_{12}:= 
\begin{bmatrix}
{\bar\chi} \\ {\bar\zeta}
\end{bmatrix}^T 
{\bar W}_{12}
\begin{bmatrix}
{\bar\chi} \\ {\bar\zeta}
\end{bmatrix},
\end{equation}
where ${\bar W}_{12}={\bar W}_{12}^T\in{\mathbb R}^{({\bar m}+{\bar n})\times ({\bar m}+{\bar n})}$ is such that $\mbox{ rank }{\bar W}_{12}={\bar m}+{\bar n}$, 
$n_-( {\bar W}_{12} )= {\bar n}$, and ${\bar W}_{12}$ satisfies the realizability condition (Assumption~\ref{AssumptionRealizability01}). Specifically, ${\bar W}_{12}$ can be obtained from ${\hat W}_{12}$ by removing the rows and the columns corresponding to those elements of ${\chi}$ and ${\zeta}$  that belong to ${\tilde\chi}$ and ${\tilde\zeta}$, respectively, and possibly recalculating the parameter $\epsilon_0\ge 0$.  

{\bf Proof:} Taking into account ${\tilde\chi}(\cdot)\equiv 0$, one can write 
\begin{equation*}
{\hat s}_{12}:= 
\begin{bmatrix}
\chi \\ \zeta
\end{bmatrix}^T 
{\hat W}_{12}
\begin{bmatrix}
\chi \\ \zeta
\end{bmatrix}
=
\begin{bmatrix}
{\bar\chi} \\ {\bar\zeta}
\end{bmatrix}^T 
{\bar W}_{12}
\begin{bmatrix}
{\bar\chi} \\ {\bar\zeta}
\end{bmatrix}-\delta\left| {\tilde\zeta}\right|^2\le \begin{bmatrix}
{\bar\chi} \\ {\bar\zeta}
\end{bmatrix}^T 
{\bar W}_{12}
\begin{bmatrix}
{\bar\chi} \\ {\bar\zeta}
\end{bmatrix}={\bar s}_{12},
\end{equation*}
where 
\begin{equation*}
{\bar W}_{12}:=\begin{bmatrix}
\theta \left({\bar W}_{\chi}+{\bar\epsilon}_0{\mathbb I}_{\bar m}\right) & \theta {\bar W}_{cross} \\
\theta {\bar W}^T_{cross} & -\delta{ \mathbb I}_{\bar n}
\end{bmatrix},
\end{equation*}
and where ${\bar W}_{\chi}$ and ${\bar W}_{cross}$ are obtained from ${W}_{\chi}$ and ${W}_{cross}$, respectively, by removing the rows and the columns that correspond to those components of input and output that belong to  ${\tilde\chi}$ and ${\tilde\zeta}$, respectively, and ${\bar\epsilon}_0\ge 0$ is chosen such that $\left({\bar W}_{\chi}+{\bar\epsilon}_0{\mathbb I}\right) + \theta\delta^{-1} {\bar W}_{cross} {\bar W}^T_{cross}$ is positive definite. It is easy to see that ${\bar W}_{12}$ has all the same properties as ${\hat W}_{12}$ does, except for specific dimensions of the blocks that the matrices comprise. The proof therefore can be completed using exactly the same line of reasoning that is used in the proof of Corollary~\ref{CorollaryDissipativitySigma12D}. $\square$

%\newpage 

\section{Example}\label{SectionExample01} In this section, a numerical example is presented that illustrates the interconnection controller design methods developed above in the theoretical part of this paper. Consider two dissipative networks $\Sigma_1$ and $\Sigma_2$. Suppose network $\Sigma_1$ has dynamics similar to that of a 2-DOF rigid mechanical system, and described by the Euler-Lagrange equations of the form 
\begin{equation}\begin{bmatrix} d_{11} & d_{12}\\
d_{21} & d_{22}
\end{bmatrix}
\begin{bmatrix} {\ddot q}_{1} \\ {\ddot q}_{2}
\end{bmatrix}
+\begin{bmatrix} c_{11} & c_{12}\\
c_{21} & 0
\end{bmatrix}
\begin{bmatrix} {\dot q}_{1} \\ {\dot q}_{2}
\end{bmatrix}+\begin{bmatrix} {\mathcal G}_{1} \\ {\mathcal G}_{2}
\end{bmatrix}= \begin{bmatrix} {\tau}_{1} \\ {\tau}_{2}
\end{bmatrix},\label{ExampleSystem01EL}
\end{equation}
where ${q}_{i}$, ${\dot q}_{i}$, and ${\ddot q}_{i}$, $i=1,2$,  are joint positions, velocities, and accelerations, respectively, $\tau_1$ and $\tau_2$ are external torques, and 
\begin{equation*}
\begin{aligned}
d_{11}&=\mu_1l^2_{c1}+\mu_2\left( l_1^2+l^2_{c2}+2l_1l_{c2}\cos q_2 \right)+I_1+I_2,\\
d_{12}& =d_{21}=\mu_2\left(l^2_{c2}+l_1l_{c2}\cos q_2  \right)+I_2,\\
d_{22}&=\mu_2l^2_{c2}+I_2,\\
c_{11}&= -\mu_2l_1 l_{c2}\sin q_2 {\dot q}_2,\\
c_{12}&=-\mu_2l_1 l_{c2}\sin q_2\left({\dot q}_1+{\dot q}_2\right),\\
c_{21}&= \mu_2l_1 l_{c2}\sin q_2 {\dot q}_1,\\
{\mathcal G}_1&=\left(\mu_1l_{c1}+\mu_2l_1  \right) g \cos q_1 + \mu_2l_{c2} g \cos(q_1+q_2),\\ 
{\mathcal G}_2&= \mu_2l_{c2} g \cos(q_1+q_2),
\end{aligned}
\end{equation*}
where $\mu_i>0$, $l_i>l_{ci}>0$, $I_i>0$, $i=1,2$, and $g>0$ are various physical parameters. In the simulations presented below, the following parameter values are used: $\mu_1=3$, 
$\mu_2=2$, $l_1=0.8$, $l_{c1}=0.4$, $l_2=0.6$, $l_{c2}=0.3$, $I_1=0.2$, $I_2=0.15$, and $g=9.81$.  It is well-known~\cite[Section 6.5.1]{spong:hutchinson:vidyasagar:2020}, that the dynamics of such a system is passive. Specifically, let ${\mathcal V}_1:={\rm K}+{\rm P}$, where 
\begin{equation*}
\begin{aligned}
{\rm K}&:=\frac{1}{2} \begin{bmatrix} {\dot q}_{1} \\ {\dot q}_{2}
\end{bmatrix}^T\begin{bmatrix} d_{11} & d_{12}\\
d_{21} & d_{22}
\end{bmatrix}
\begin{bmatrix} {\dot q}_{1} \\ {\dot q}_{2}
\end{bmatrix},\\
{\rm P}&:= \left(\mu_1l_{c_1}+ \mu_2l_{1}\right)g\sin q_1 + \mu_2l_{c_2}g\sin (q_1+q_2)
\end{aligned}
\end{equation*}
%and 
%\begin{equation*}
%{\rm P}:= \left(m_1l_{c_1}+ m_2l_{1}\right)g\sin q_1 + m_2l_{c_2}g\sin (q_1+q_2) \end{equation*}
 are the kinetic and the potential energy of the system~\eqref{ExampleSystem01EL}, respectively. 
Denote 
\begin{equation*}
\eta_1=\begin{bmatrix} {\tau}_{1} \\ {\tau}_{2}
\end{bmatrix}, \hspace{0.3cm} 
y_1=\begin{bmatrix} {\dot q}_{1} \\ {\dot q}_{2}
\end{bmatrix}. 
\end{equation*}
Then ${\dot {\mathcal V}}_1=s_1$, where 
\begin{equation}\label{System1passivity}
s_1=\frac{1}{2}\begin{bmatrix}
\eta_1\\ y_1
\end{bmatrix}^T
\begin{bmatrix}
0 & {\mathbb I}_2\\
{\mathbb I}_2 & 0
\end{bmatrix}
\begin{bmatrix}
\eta_1\\ y_1
\end{bmatrix}.
\end{equation} 

On the other hand, suppose $\Sigma_2$ is described by the following equations: 
\begin{equation*}
\begin{aligned}
{\dot x}_2^{\{1\}} &= -\left(3-e^{-\left(x_2^{\{2\}}\right)^2}\right)x_2^{\{1\}}+\eta_2,\\
{\dot x}_2^{\{2\}} &= x_2^{\{1\}}-x_2^{\{2\}},
\end{aligned}
\end{equation*}
where 
\begin{equation*}
y_2=\begin{bmatrix}y_2^{\{1\}}\\ y_2^{\{2\}} \end{bmatrix}:=\begin{bmatrix}x_2^{\{1\}}\\ x_2^{\{2\}} \end{bmatrix}
\end{equation*}
The system $\Sigma_2$ is dissipative with a supply rate of the form
\begin{equation*} 
s_2=
\begin{bmatrix}\eta_2\\y_2^{\{1\}}\\ y_2^{\{2\}} \end{bmatrix}^T
\begin{bmatrix}
  0  &  0.5       &  0\\
    0.5 & -2 &   0.5\\
         0  &  0.5 & -1
\end{bmatrix}
\begin{bmatrix}\eta_2\\y_2^{\{1\}}\\ y_2^{\{2\}} \end{bmatrix}. 
\end{equation*} 
Indeed, pick a storage function candidate $V_2=\frac{1}{2}|x_2^{\{1\}}|^2 + \frac{1}{2}|x_2^{\{2\}}|^2$. Taking into account that $e^{-\left(x_2^{\{2\}}\right)^2}\le 1$, the time derivative of $V_2$ along the trajectories of $\Sigma_2$ is 
\begin{equation*}
{\dot V}_2=x_2^{\{1\}}{\dot x}_2^{\{1\}}+x_2^{\{2\}}{\dot x}_2^{\{2\}}\le -2|x_2^{\{1\}}|^2+x_2^{\{1\}}\eta_2+x_2^{\{1\}}x_2^{\{2\}}-|x_2^{\{2\}}|^2=s_2. 
\end{equation*}

\begin{figure}[h!]
    \begin{center}
        \includegraphics[scale=0.6]{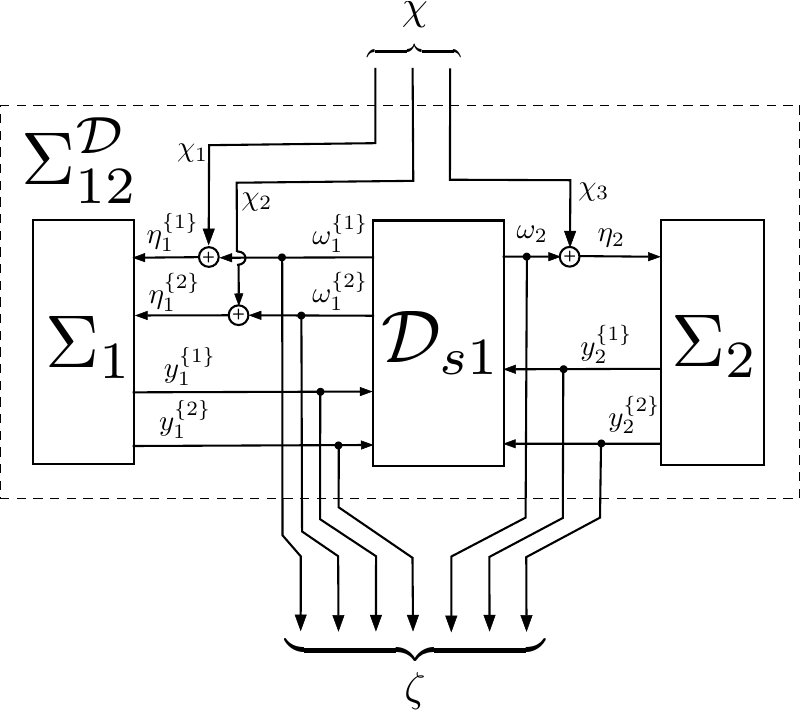}
       \caption{Interconnection $\Sigma_{12}^{\mathcal D}$}
        \label{Fig:FigureExample02Step01a}
    \end{center}
\end{figure}

The interconnection $\Sigma_{12}^{\mathcal D}$ which comprises $\Sigma_1$, $\Sigma_2$,  and a dynamic controller ${\mathcal D}_{s1}$ is shown in Figure~\ref{Fig:FigureExample02Step01a}. 
Our goal is to design the dynamic scattering controller ${\mathcal D}_{s1}$ according to the method developed in the theoretical part of this paper. MATLAB script that performs all calculations can be found in~\cite{PolushinGithubMATLABDynamicScatteringControllerDesignNov4thmMain}. We have $m_1=2$, $p_1=2$, $m_2=1$, $p_2=2$, and therefore ${\mathcal N}_1=m_1+p_1=4$, ${\mathcal N}_2=m_2+p_2=3$, ${\mathfrak m}:={\mathfrak m}=3$, ${\mathfrak p}:=p_1+p_2=4$, and 
${\mathfrak n}:=\max\{ {\mathfrak m}, {\mathfrak p}\}=4$. The free design parameters were chosen as follows. The permutation matrices $P_1$, $P_2$ in~\eqref{Pi01} are chosen to be unit matrices of the corresponding dimensions. Also, matrices involved in~\eqref{LyuapunovEquation01}, \eqref{W0} are chosen as follows $A_0=-{\mathbb I}_4$, $B_0={\mathbb I}_4$, and ${\mathcal Q}={\mathbb I}_4$. Furthermore, $\delta=0.5$, $\Lambda_*^+={\mathbb I}_4$, and $\theta=3$; $\Lambda_{\bf u}$ is subsequently calculated using~\eqref{Ineq02a}, where the equal sign is used instead of $\succeq$. We choose ${\mathcal W}$ in~\eqref{WeightingMatrixW01} to be $3\times 4$ matrix where every element is equal to $0.25$, calculated that $\sigma_{\max}\left( {\mathcal W}^T\Lambda_{\bf u} {\mathcal W}\right)=1.2232$, and consequently choose $\Lambda_*^-=1.2232\cdot {\mathbb I}_4$. Finally, matrix ${\rm P}_d={\mathbb I}$ in~\eqref{PD01}. 

Calculations performed by the MATLAB script yield the following results. The scattering transformations ${\mathbb S}^\sharp_i$, ${\mathbb S}^\sharp_i$ in~\eqref{ScatteringVariablesSharpForS1S2} are equal to 
\begin{equation*} 
{\mathbb S}^\sharp_1=\begin{bmatrix}
         0  & -{\sqrt 2}&  -1   &     0\\
    {\sqrt 2}    &     0     &    0  & -1\\
         0  &  1  &  {\sqrt 2}      &   0\\
    1  &     0     &    0  & -{\sqrt 2}
\end{bmatrix}, \hspace{0.3cm}
%\end{equation*}
%\begin{equation*}
{\mathbb S}^\sharp_2=\begin{bmatrix}
 1.0402  & -0.2618 &  -0.1157\\
    0.1979 &  -0.3657  & -0.9515\\
    0.2068  & -0.9668  &  0.3286
\end{bmatrix}.
\end{equation*}
Matrices involved in the controller~\eqref{ControllerD01} are 
\begin{equation*}
A=-\sqrt{2}\cdot {\mathbb I}_4, \hspace{0.5cm} B=-2.3784\cdot {\mathbb I}_4
\end{equation*}
\begin{equation*}
C=1.0753\cdot 
\begin{bmatrix}
         1  & 0 &  0   &     0\\
    0  &     0    &    1  & 0\\
         0  &  1  &  0     &   0\\
    0  &     0     &    0  & 1
\end{bmatrix}, \hspace{0.5cm} 
D=0.9042\cdot 
\begin{bmatrix}
         1  & 0 &  0   &     0\\
    0  &     0    &    1  & 0\\
         0  &  1  &  0     &   0\\
    0  &     0     &    0  & 1
\end{bmatrix}
\end{equation*}
The matrices of interconnection supply rates in~\eqref{s12Definition} and~\eqref{SupplyRateHatS12} are 
\begin{equation*}
W_{12}={\hat W}_{12}=
\begin{bmatrix}0 & 0 & 0 & 0 & 0 & 1.5 & 0 & 0 & 0 & 0\\ 0 & 0 & 0 & 0 & 0 & 0 & 1.5 & 0 & 0 & 0\\ 0 & 0 & 0 & 0 & 0 & 0 & 0 & 0 & 1.5 & 0\\
 0 & 0 & 0 & -1 & 0 & 0 & 0 & 0 & 0 & 0\\ 0 & 0 & 0 & 0 & -1 & 0 & 0 & 0 & 0 & 0\\ 1.5 & 0 & 0 & 0 & 0 & -1 & 0 & 0 & 0 & 0\\ 
 0 & 1.5 & 0 & 0 & 0 & 0 & -1 & 0 & 0 & 0\\ 0 & 0 & 0 & 0 & 0 & 0 & 0 & -1 & 0 & 0\\ 0 & 0 &1.5 & 0 & 0 & 0 & 0 & 0 & -1 & 0\\ 0 & 0 & 0 & 0 & 0 & 0 & 0 & 0 & 0 & -1 
\end{bmatrix}
\end{equation*} 
and matrix ${\mathcal L}_2$-gain in~\eqref{InequalityL2GainOfSigmaD12} is 
\begin{equation*}
\Upsilon_{12}=36\cdot {\mathbb I}_3. 
\end{equation*}

\begin{figure}[h!]
    \begin{center}
        \includegraphics[scale=0.4]{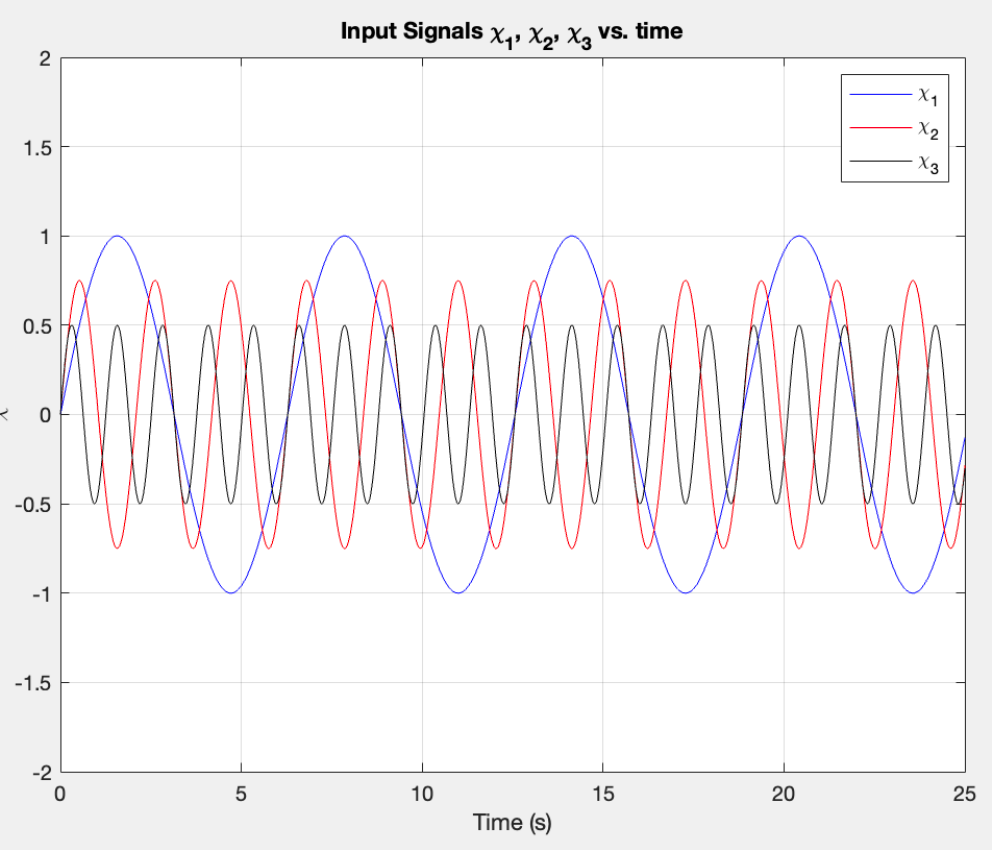}
       \caption{Response of $\Sigma_{12}^{\mathcal D}$ to sinusoidal excitation signals: input signals $\chi_1$, $\chi_2$, $\chi_3$ vs. time}
        \label{Fig:FigureExperiment1Plot1}
    \end{center}
\end{figure}

\begin{figure}[h!]
    \begin{center}
        \includegraphics[scale=0.4]{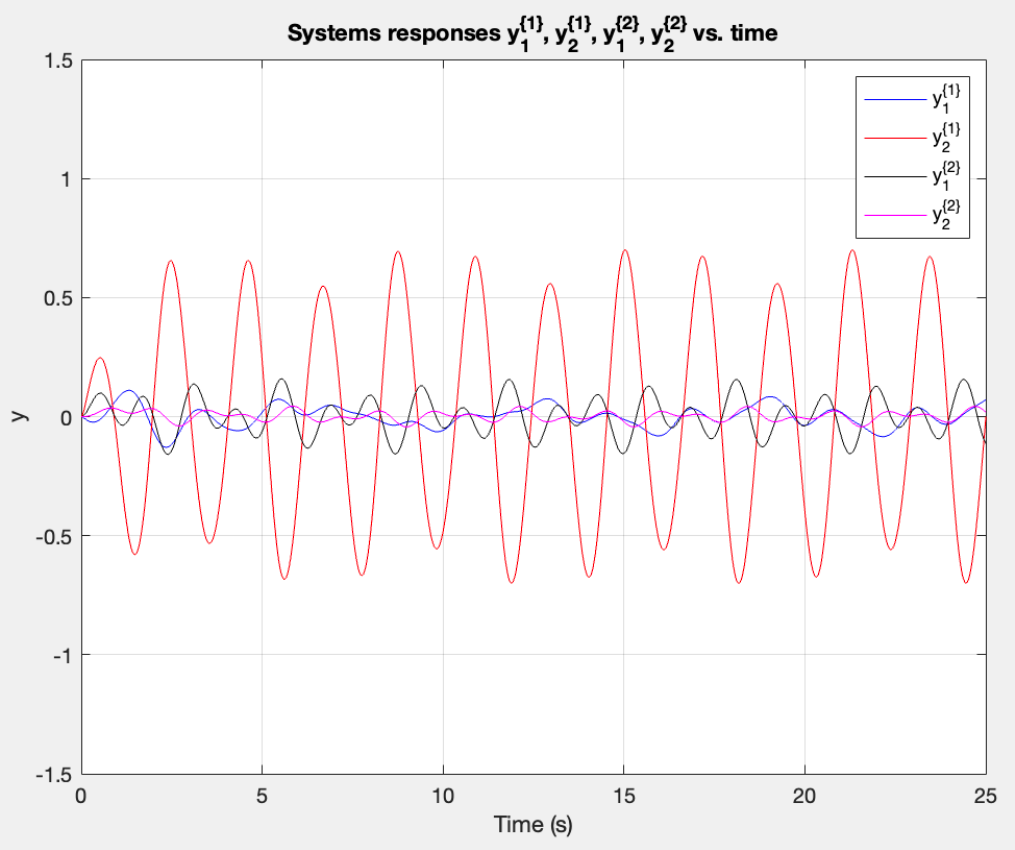}
       \caption{Response of $\Sigma_{12}^{\mathcal D}$ to sinusoidal excitation signals: systems' responses $y_1^{\{1\}}$, $y_1^{\{2\}}$, $y_2^{\{1\}}$, $y_2^{\{2\}}$ vs. time}, 
        \label{Fig:FigureExperiment1Plot2}
    \end{center}
\end{figure}

\begin{figure}[h!]
    \begin{center}
        \includegraphics[scale=0.4]{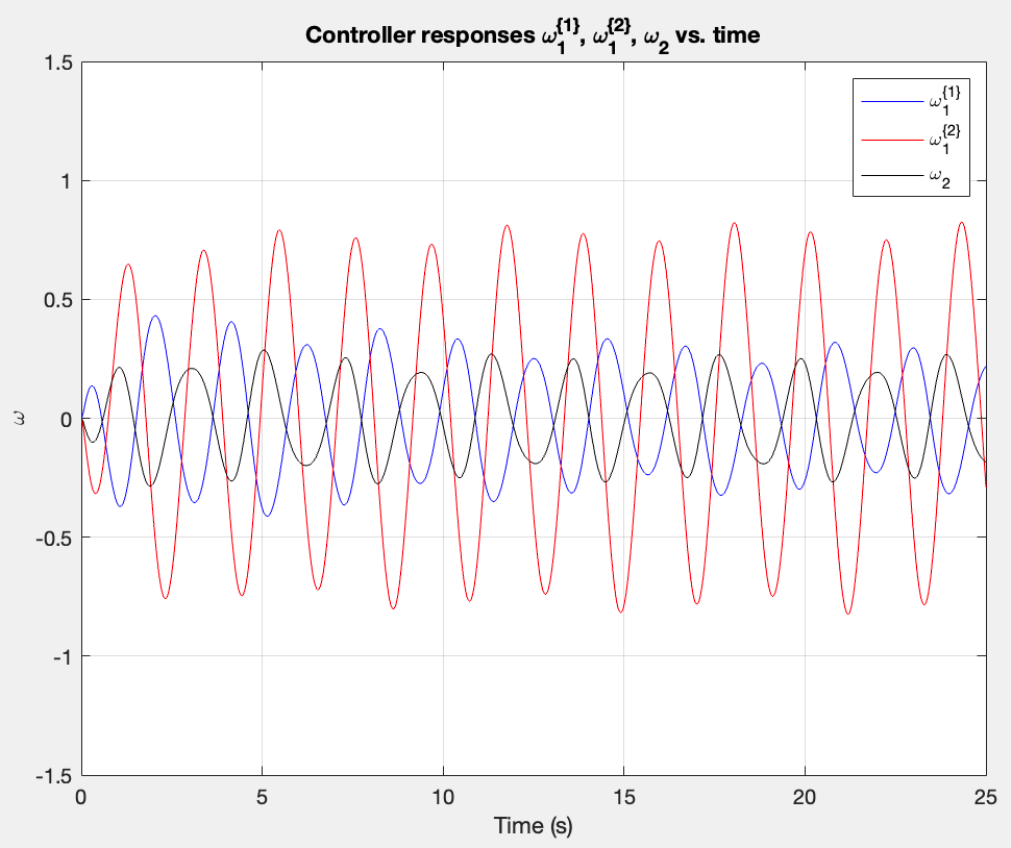}
       \caption{Response of $\Sigma_{12}^{\mathcal D}$ to sinusoidal excitation signals: controller response $\omega_1^{\{1\}}$, $\omega_1^{\{2\}}$, $\omega_2$ vs. time}
        \label{Fig:FigureExperiment1Plot3}
    \end{center}
\end{figure}

\begin{figure}[h!]
    \begin{center}
        \includegraphics[scale=0.4]{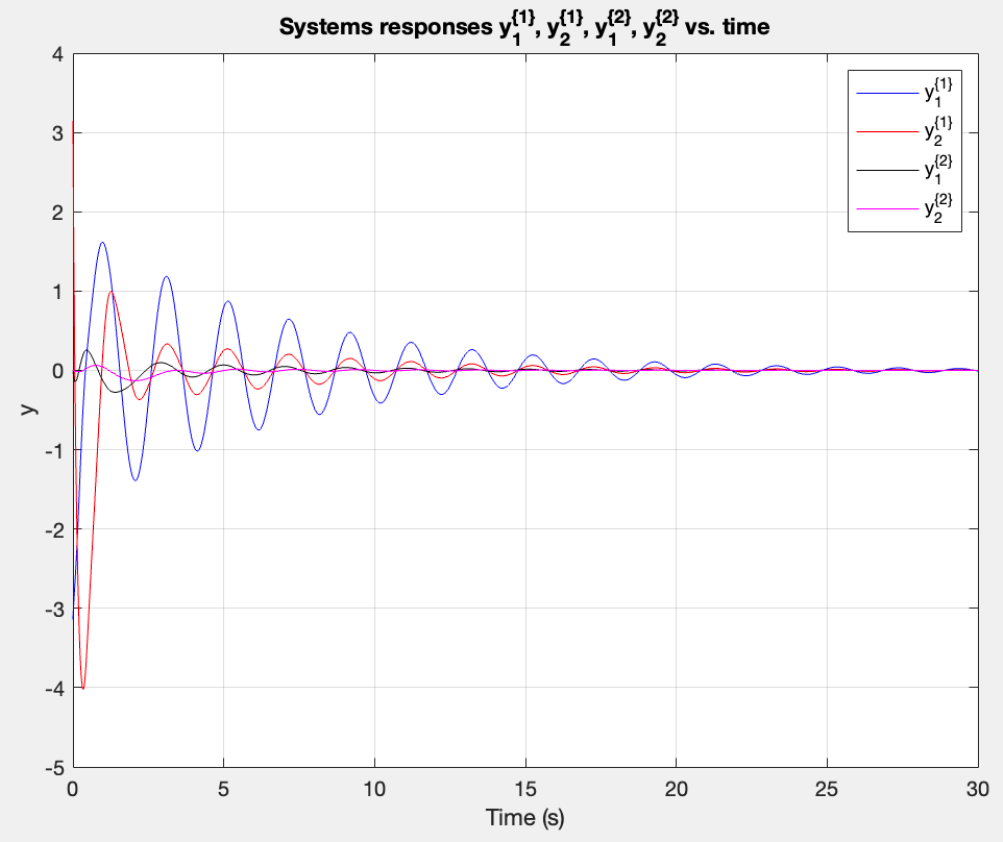}
       \caption{Response of $\Sigma_{12}^{\mathcal D}$ to non-equilibrium initial conditions: systems' responses $y_1^{\{1\}}$, $y_1^{\{2\}}$, $y_2^{\{1\}}$, $y_2^{\{2\}}$ vs. time}
        \label{Fig:FigureExperiment3Plot2}
    \end{center}
\end{figure}

\begin{figure}[h!]
    \begin{center}
        \includegraphics[scale=0.4]{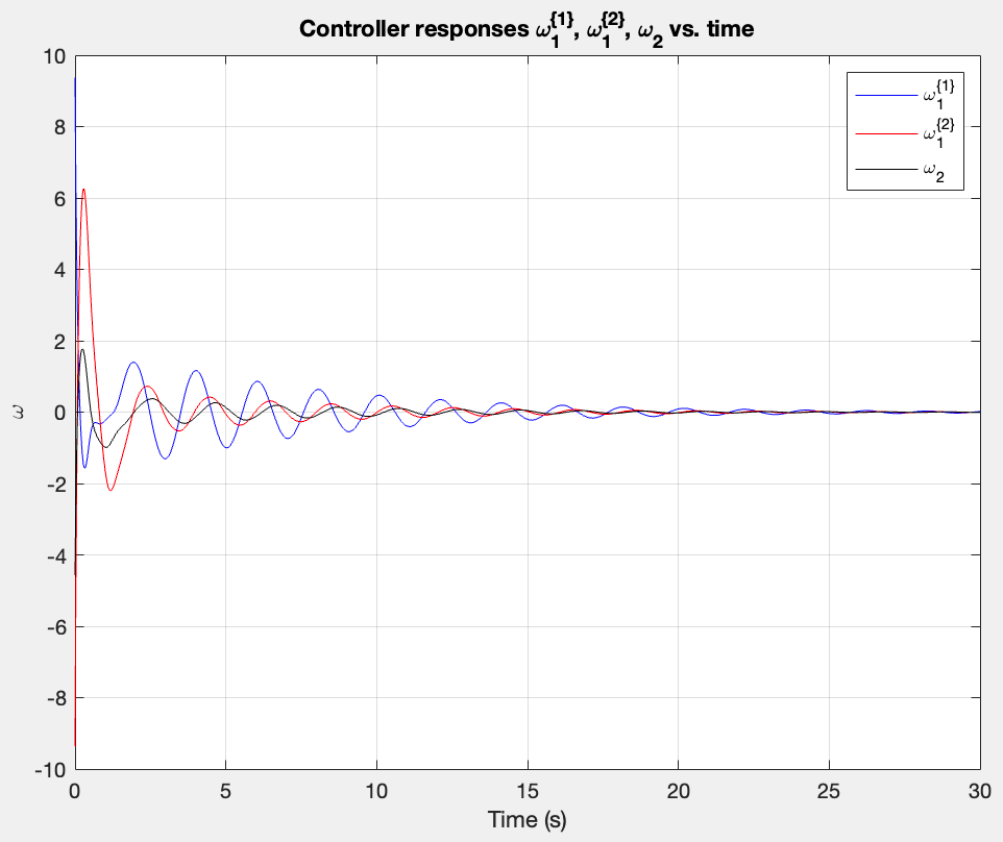}
       \caption{Response of $\Sigma_{12}^{\mathcal D}$ to non-equilibrium initial conditions: controller response $\omega_1^{\{1\}}$, $\omega_1^{\{2\}}$, $\omega_2$ vs. time}
        \label{Fig:FigureExperiment3Plot3}
    \end{center}
\end{figure}

{MATLAB}   simulations were performed that demonstrate responses of the interconnection $\Sigma_{12}^{\mathcal D}$ to external signals $\chi_1$, $\chi_2$, $\chi_3$ as well as to non-equilibrium initial conditions. Simulation results are illustrated in Figures~\ref{Fig:FigureExperiment1Plot1}-\ref{Fig:FigureExperiment3Plot3}. Specifically, Figures~\ref{Fig:FigureExperiment1Plot1}-\ref{Fig:FigureExperiment1Plot3} demonstrate response of the interconnection $\Sigma_{12}^{\mathcal D}$ to sinusoidal excitation signals $\chi_1(t)=\cos t$, $\chi_2(t)=0.75\cos 3t$, $\chi_3=0.5 \cos 5t$, assuming initial conditions for all subsystems are at their corresponding equilibrium points ($q_1(0)=-\pi/2$, $q_2(0)=0$, ${\dot q}_1(0)=0$, ${\dot q}_2(0)=0$, $x_2^{\{1\}}(0)=0$,  $x_2^{\{2\}}(0)=0$, and all controller state variables are initially equal to zero). Figure~\ref{Fig:FigureExperiment1Plot2} demonstrates the response of systems' variables $y_1^{\{1\}}$, $y_1^{\{2\}}$, $y_2^{\{1\}}$, $y_2^{\{2\}}$, while Figure~\ref{Fig:FigureExperiment1Plot3} shows the response of controller variables $\omega_1^{\{1\}}$, $\omega_1^{\{2\}}$, and $\omega_2$. As can be seen from the plots, the interconnection demonstrates bounded response to external sinusoidal excitation signals. 

Response of the $\Sigma_{12}^{\mathcal D}$ to non-equilibrium initial conditions are shown in Figures~\ref{Fig:FigureExperiment3Plot2} and~\ref{Fig:FigureExperiment3Plot3}. Specifically, these figures corresponds to the case where the initial conditions of subsystem $\Sigma_1$ are as follows: $q_1(0)=-\pi/2$, $q_2(0)=\pi/2$, ${\dot q}_1(0)=-\pi$, ${\dot q}_2(0)=\pi$, all other initial conditions are zero. External excitation signals $\chi_1$, $\chi_2$, $\chi_3$ are all totally equal to zeros. Response of systems' variables $y_1^{\{1\}}$, $y_1^{\{2\}}$, $y_2^{\{1\}}$, $y_2^{\{2\}}$ are shown in Figure~\ref{Fig:FigureExperiment3Plot2}, while those of the controller variables $\omega_1^{\{1\}}$, $\omega_1^{\{2\}}$, and $\omega_2$ in Figure~\ref{Fig:FigureExperiment3Plot3}. It can be seen that, in the absence of external excitation, all interconnection variables converges to zeros. 

%\newpage 
 
\section{Conclusions}\label{SectionConclusions01} In this paper, the problem of stable interconnection of dissimilar dissipative networks through a dynamic scattering controller is addressed. Under basic assumptions of dissipativity with a quadratic supply rate, liveness, and realizability imposed on the subsystems, design of dynamic scattering controller is presented that guarantees the interconnection to be internally stable and possesses a well-defined quadratic  supply rate. One of the major improvements brought by this work as compared to~\cite{polushin:26:automatica}, besides the introduction of a dynamic scattering controller,  is the fact that the interconnection of subsystems $\Sigma_1$ and $\Sigma_2$ is guaranteed to satisfy  the same basic assumptions as those originally imposed on the subsystems. This allows such an interconnection to be further interconnected with other dissipative networks with similar properties using exactly the same method. Because the basic assumptions are guaranteed to hold at every step,  the process can be repeated an arbitrary number of times which allows for building networks of dissipative systems of arbitrary large scale. Applications of the developed techniques to practical problems, such as power systems,  that involve large scale interconnections of dissipative networks is of  interest for future research.

%\bibliographystyle{IEEEtran}
%\bibliography{/Users/ilia/Documents/Bibliography/base}

%\newpage 

\appendices

\section{Proofs}\label{AppendixProofs01}
\subsection{Proof of Lemma~\ref{LemmaA0B0}} Calculating the Schur complement of $-{\mathcal Q}$ in $W_0$, one gets 
\begin{equation*}
\begin{bmatrix}
{\mathbb I} & B_0^T{\mathcal P}{\mathcal Q}^{-1}\\
{\mathbb O} & {\mathbb I}
\end{bmatrix}
\begin{bmatrix}
{\mathbb O} & B_0^T{\mathcal P}\\
{\mathcal P}B_0 & -{\mathcal Q}
\end{bmatrix}
\begin{bmatrix}
{\mathbb I} & {\mathbb O} \\
 {\mathcal Q}^{-1}{\mathcal P}B_0 & {\mathbb I}
\end{bmatrix}
=\begin{bmatrix}
B_0^TPQ^{-1}{\mathcal P}B_0 & {\mathbb O} \\
{\mathbb O}  & -{\mathcal Q}
\end{bmatrix}.
\end{equation*}
The latter matrix is block diagonal, where the lower-right block (of the size ${\mathfrak n}\times {\mathfrak n}$) is negative definite while the upper-left block (of the size ${\mathfrak p}\times {\mathfrak p}$) is positive definite, the latter follows from the facts that ${\mathcal P}{\mathcal Q}^{-1}{\mathcal P}$ is positive definite and $B_0$ has full rank. 
By Sylvester's criteria~\cite[Theorem 4.5.8]{horn:johnson:13}, the statement of Lemma~\ref{LemmaA0B0} follows.

\subsection{Proof of Lemma~\ref{LemmaRankG22}}\label{ProofLemmaG22Rank} Denote
\begin{equation*}
G_{-}:=\begin{bmatrix}
{G}_{12} \\
{G}_{22} 
\end{bmatrix}. 
\end{equation*}
By definition, the columns of $G_{-}$ are orthonormal eigenvectors that correspond to negative eigenvalues of $W_0$, {\it i.e.}
\begin{equation*}
G_{-}^TW_0G_{-}=-\Lambda^D_- 
\end{equation*}
Denote ${\mathfrak M}_1:=\{ x\in{\mathbb R}^{{\mathfrak n}+{\mathfrak p}}\vert\, x_{{\mathfrak p}+1}=\ldots x_{{\mathfrak p}+{\mathfrak n}}=0\}$, and 
${\mathfrak M}_2:=\{ x\in{\mathbb R}^{{\mathfrak n}+{\mathfrak p}}\vert\, x_{1}=\ldots x_{{\mathfrak n}}=0\}$. Clearly, ${\mathfrak M}_1$ and ${\mathfrak M}_2$ are complementary subspaces of ${\mathbb R}^{{\mathfrak n}+{\mathfrak p}}$, {\it i.e.}, ${\mathfrak M}_1 \cap {\mathfrak M}_2=\{ 0 \}$ and ${\mathfrak M}_1 \oplus {\mathfrak M}_2={\mathbb R}^{{\mathfrak n}+{\mathfrak p}}$. Also, let $S_-\subset {\mathbb R}^{{\mathfrak n}+{\mathfrak p}}$ be a subspace spanned by the columns of $G_{-}$. 

First, let us show that ${\mathfrak M}_1 \cap S_-=\{ 0 \}$. Let $x\in {\mathfrak M}_1 \cap S_-$. By Rayleigh quotient theorem~\cite[Theorem 4.2.2]{horn:johnson:13}, for any nonzero $x\in S_-$ one has 
$x^T W_0x<0$. On the other for any $x\in {\mathfrak M}_1$ we have 
\begin{equation*}
x^TW_0x=
\begin{bmatrix}
x_1\\ \vdots \\ x_{\mathfrak p}\\ 0 \\ \vdots \\ 0
\end{bmatrix}^T
\begin{bmatrix}
{\mathbb O} & B_0^TP\\
PB_0 & -Q
\end{bmatrix}
\begin{bmatrix}
x_1\\ \vdots \\ x_{\mathfrak p}\\ 0 \\ \vdots \\ 0
\end{bmatrix}=0. 
\end{equation*}
Thus, ${\mathfrak M}_1 \cap S_-=\{ 0 \}$. 

Now, let $E_2\colon\,{\mathbb R}^{{\mathfrak n}+{\mathfrak p}}\to {\mathfrak M}_2$ be a projection operator on ${\mathfrak M}_2$, and 
${\tilde E}_2:=E_2\vert_{S_-}$ be a restriction of  ${\rm Proj}_{{\mathfrak M}_2}$ onto the subspace $S_-$, {\it i.e.}, ${\tilde E}_2(x)=E_2(x)$ for any $x\in S_-$. We state that ${\tilde E}_2$ is full rank and therefore bijective. Indeed, let $x\in S_-$ be such that ${\tilde E}_2(x)=0$. This implies that $x\in {\mathfrak M}_1$, and therefore $x=0$. Thus, nullity (dimension of the kernel) of ${\tilde E}_2$ is equal to $0$, and  ${\tilde E}_2$ is full rank. Since dimensions of $S_-$ and ${\mathfrak M}_2$ are both equal to ${\mathfrak n}$, ${\tilde E}_2$ is bijective. 

Now, the columns of $G_{-}$ are orthonormal and therefore form a basis of $S_-$. Projection operator ${\tilde E}_2$ maps each of these column vectors to the corresponding column of $G_{22}$. Since a linear bijective operator  maps a basis into a basis, it implies that columns of $G_{22}$ form a basis of ${\mathfrak M}_2$. The statement of Lemma~\ref{LemmaRankG22} follows. 

%\newpage 
%
%
%\input{SectionAppendixSupplyRateAssignment01}

%\newpage 

\section{On Realizability of Scattering Transformations}\label{AppendixRealizability01}
\begin{figure}[h!]
    \begin{center}
        \includegraphics[scale=0.6]{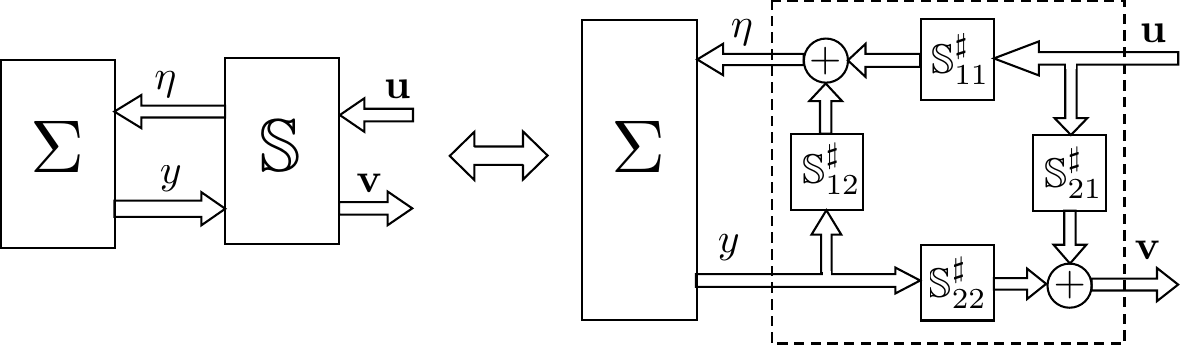}
       \caption{Scattering transformation}
        \label{Fig:Figure09}
    \end{center}
\end{figure}
The scattering transformation addressed in this as well as previous~\cite{usova:polushin:patel:18:automatica,usova:polushin:patel:19:cssletters,polushin:26:automatica} works is a linear input-output transformation of the form
\begin{equation}\label{ScatteringTransformationApp1}
\begin{bmatrix}
\eta\\ y
\end{bmatrix}=
{\mathbb S}
\begin{bmatrix}
{\bf u}\\ {\bf v}
\end{bmatrix}
\end{equation}
where $(\eta , y)$ and $({\bf u}, {\bf v})$ are the original and the transformed input-output pairs, respectively, $\eta, {\bf u}\in {\mathbb R}^{m}$, $y, {\bf v}\in {\mathbb R}^{p}$. When the scattering transformation ${\mathbb S}$ is implemented as a part of an interconnection, as shown in Figure~\ref{Fig:Figure09}, signals ${\bf u}$ and $y$ are 
%determined by 
essentially outputs of other subsystems involved in the interconnection and therefore play the role of inputs to the scattering transformation, while $\eta$ and ${\bf v}$ are the outputs of ${\mathbb S}$ which are  calculated as functions of ${\bf u}$ and $y$. In other words, in order to successfully implement the scattering transformation, one needs to find a transformation ${\mathbb S}^\sharp$ of the form
\begin{equation}\label{Ssharp01}
\begin{bmatrix}
\eta\\ {\bf v}
\end{bmatrix}=
{\mathbb S}^\sharp
\begin{bmatrix}
{\bf u}\\ y
\end{bmatrix}
=\begin{bmatrix}
{\mathbb S}^\sharp_{11} & {\mathbb S}^\sharp_{12}\\
{\mathbb S}^\sharp_{21} & {\mathbb S}^\sharp_{22}
\end{bmatrix} 
\begin{bmatrix}
{\bf u}\\ y
\end{bmatrix}
\end{equation}
which is equivalent to ${\mathbb S}$ in the sense that it defines an equivalent relation between the signals, see also Figure~\ref{Fig:Figure09} (right). 

Suppose the system $\Sigma$ is dissipative with a quadratic supply rate of the form
\begin{equation}\label{SupplyRateAppendix01}
s=\begin{bmatrix}
\eta\\ y
\end{bmatrix}^T
W
\begin{bmatrix}
\eta\\ y
\end{bmatrix}
\end{equation}
where $W=W^T\in{\mathbb R}^{({m+p})\times ({m+p})}$. A general form of the scattering transformation for this system is 
\begin{equation}\label{FormulaScatteringTrans01}
{\mathbb S}:={G} {\mathcal P}\, \Gamma.
\end{equation} 
In the above formula~\eqref{FormulaScatteringTrans01}, $G\in{\mathbb R}^{({m+p})\times ({m+p})}$ is an orthonormal matrix that satisfy equation
\begin{equation}\label{Decomposition01}
WG=\Lambda G,
\end{equation}
where $\Lambda\in{\mathbb R}^{({m+p})\times ({m+p})}$ is a diagonal matrix whose diagonal elements are the eigenvalues $W$ written in the descending order. Also, ${\mathcal P}\in{\mathbb R}^{({m+p})\times ({m+p})}$ is a permutation matrix of the form
\begin{equation}\label{FormulaPermMatrix01}
{\mathcal P}:=
\begin{bmatrix}
{\mathcal P}_+ & {\mathbb O}\\
{\mathbb O} & {\mathcal P}_- 
\end{bmatrix},
\end{equation}
where ${\mathcal P}_+\in {\mathbb R}^{{m}\times {m}}$, ${\mathcal P}_-\in {\mathbb R}^{{p}\times {p}}$, and 
$\Gamma\in{\mathbb R}^{({m+p})\times ({m+p})}$ is a diagonal matrix of positive gains, $\Gamma\succ 0$. 

Sufficient condition for realizability of the scattering transformation used in this paper can be formulated as follows. 
Let ${\mathbb W}$ be a signal space which consists of all possible values of manifest variables $w:=(\eta, y)$, and let ${\mathcal Y}:=\left\{w\in{\mathbb W}\vert \, \eta=0\right\}$. Also, let ${\mathcal G}_{\{-\}}$ be a subspace spanned by $p$ eigenvectors of the supply rate matrix $W$ that correspond to $p$ smallest eigenvalues of $W$. Sufficient condition for realizability is that projection of ${\mathcal G}_{\{-\}}$ onto ${\mathcal Y}$ is bijective. In the specific coordinates determined by~\eqref{Decomposition01}, the above condition is obviously equivalent to the following:
\begin{equation}\label{ConditionRankG22a}
\mbox{rank } G_{22}=p,
\end{equation}
where $G_{22}\in{\mathbb R}^{p\times p}$ is the bottom right block of the matrix $G$ defined in~\eqref{Decomposition01}, specifically:
\begin{equation}\label{FormulaForG01}
G=\begin{bmatrix}
G_{11} & G_{12}\\
G_{21} & G_{22}
\end{bmatrix} 
\end{equation}
where the remaining blocks are $G_{11}\in{\mathbb R}^{m\times m}$, and $G_{12}$, $G_{21}^T\in {\mathbb R}^{m\times p}$. 
Under condition~\eqref{ConditionRankG22a}, the submatrices of ${\mathbb S}^\sharp$ in~\eqref{Ssharp01} can be calculated as follows. Using~\eqref{FormulaPermMatrix01} and~\eqref{FormulaForG01}, formula~\eqref{FormulaScatteringTrans01} can be rewritten in the form
\begin{equation*}
{\mathbb S}=\begin{bmatrix}
{\mathbb S}_{11} & {\mathbb S}_{12}\\
{\mathbb S}_{21} & {\mathbb S}_{22}
\end{bmatrix} =\begin{bmatrix}
G_{11}{\mathcal P}_+\Gamma_+ & G_{12}{\mathcal P}_-\Gamma_-\\
G_{21}{\mathcal P}_+\Gamma_+ & G_{22}{\mathcal P}_-\Gamma_-
\end{bmatrix}  
\end{equation*} 
where $\Gamma_+\in{\mathbb R}^{{m}\times {m}}$, $\Gamma_-\in{\mathbb R}^{{p}\times {p}}$ are diagonal positive definite matrices that comprise 
$\Gamma:=\mbox{diag }\left\{\Gamma_+, \, \Gamma_-\right\}$. Submatrix ${\mathbb S}_{22}=G_{22}{\mathcal P}_-\Gamma_-$ is full rank as a product of full rank matrices. Direct calculations then show that
\begin{equation*}
\begin{aligned}
{\mathbb S}^\sharp_{11}&={\mathbb S}_{11}-{\mathbb S}_{12}{\mathbb S}^{-1}_{22}{\mathbb S}_{21}
= \left(G_{11}- G_{12}G^{-1}_{22}G_{21}\right){\mathcal P}_+\Gamma_+\\
{\mathbb S}^\sharp_{12}&={\mathbb S}_{12}{\mathbb S}^{-1}_{22}
=G_{12}{\mathcal P}_-\Gamma_-\left(\Gamma^{-1}_-{\mathcal P}^T_-G^{-1}_{22}\right)
=G_{12}G^{-1}_{22}\\
{\mathbb S}^\sharp_{21}&=-{\mathbb S}^{-1}_{22}{\mathbb S}_{21}
=-\Gamma^{-1}_-{\mathcal P}^T_-G^{-1}_{22}G_{21}{\mathcal P}_+\Gamma_+\\
{\mathbb S}^\sharp_{22}&={\mathbb S}^{-1}_{22}
=\Gamma^{-1}_-{\mathcal P}^T_-G^{-1}_{22}
\end{aligned}
\end{equation*}

\end{document}